\documentclass[submission,copyright,creativecommons]{eptcs}
\providecommand{\event}{ICLP 2025} 

\usepackage{iftex}

\usepackage{times}
\usepackage{url}
\usepackage{hyperref}
\usepackage{graphicx}
\usepackage[small]{caption}
\usepackage{amsmath}
\usepackage{amsthm}
\usepackage{amssymb}
\usepackage{bm}
\usepackage{algorithm}
\usepackage{algorithmic}
\usepackage[switch]{lineno}
\usepackage{xcolor}
\usepackage{colortbl}
\usepackage{booktabs}
\usepackage{subcaption}
\usepackage{listings}
\usepackage{bm}
\usepackage[english]{babel}

\newtheorem{example}{Example}

\newtheorem{definition}{Definition}

\ifpdf
  \usepackage{underscore}         
  \usepackage[T1]{fontenc}        
\else
  \usepackage{breakurl}           
\fi

\title{The Combined Problem of Online Task Assignment\\ and Lifelong Path Finding in Logistics Warehouses:\\ Rule-Based Systems Matter}

\author{Fengming Zhu$^1$ \quad Weijia Xu$^2$ \quad Yifei Guo$^2$ \quad Fangzhen Lin$^1$
\institute{
$^1$CSE Department,
Hong Kong University of Science and Technology, Hong Kong SAR, China\\
$^2$Meituan Academy of Robotics Shenzhen,
Shenzhen, China\\
}
\email{fzhuae@connect.ust.hk{\rm }\quad \{xuweijia, guoyifei02\}@meituan.com{\rm }\quad flin@cs.ust.hk}\\
}

\def\titlerunning{Combining Online Task Assignment and Lifelong Path Finding}
\def\authorrunning{Fengming Zhu, Weijia Xu, Yifei Guo \& Fangzhen Lin}
\begin{document}
\maketitle

\begin{abstract}
We study the combined problem of online task assignment and lifelong path finding, which is crucial for the logistics industries.
However, most literature either (1) focuses on lifelong path finding assuming a given task assigner, or (2) studies the offline version of this problem where tasks are known in advance.
We argue that, to maximize the system throughput, the online version that integrates these two components should be tackled directly.
To this end, we introduce a formal framework of the combined problem and its solution concept.
Then, we design a rule-based lifelong planner under a practical robot model that works well even in environments with severe local congestion.
Upon that, we automate the search for the task assigner with respect to the underlying path planner.
Simulation experiments conducted in warehouse scenarios at \textit{Meituan}, one of the largest shopping platforms in China, demonstrate that
(a)~\textit{in terms of time efficiency},
our system requires only 83.77\% of the execution time needed for the currently deployed system at Meituan, outperforming other SOTA algorithms by 8.09\%;
(b)~\textit{in terms of economic efficiency},
ours can achieve the same throughput with only 60\% of the agents currently in use.
The code and demos are available at \url{https://github.com/Fernadoo/Online-TAPF}.
\end{abstract}

\section{Introduction}


We consider the problem present in highly automated real-world warehouses where a fleet of robots is programmed to pick up and deliver packages without any collision.
This is a significant problem for logistics companies as it has a major impact on their operational efficiency.  
It is a difficult problem for at least the following  two reasons:
(1)~the computational complexity of multi-agent path finding is notoriously high, especially when the number of robots is large, and
(2)~the dynamic and real-time assignment of tasks to the robots both depends on and affects the subsequent path finding.

There is a vast literature that studies idealized abstractions of such real-world problems.
The most commonly seen formulation is to assume a given (or naive) task assigner, and therefore, the focus is merely on the path-finding part, which is usually termed as one-shot \textit{Multi-Agent Path Finding} (MAPF)~\cite{yu2013structure,erdem2013general,sharon2015conflict,li2021eecbs,okumura2022priority}
or its lifelong version \textit{Multi-Agent Pickup and Delivery} (MAPD)~\cite{ma2017lifelong,vsvancara2019online,li2021lifelong,okumura2022priority}.
However, to maximize the throughput of the whole production pipeline, the task assigner should also be deliberately designed with respect to the particular underlying path planner.
To this end, some recent work has further investigated the combined problem of \textit{Task Assignment and Path Finding} (TAPF)~\cite{yu2013multi,ma2016optimal,honig2018conflict,liu2019task,chen2021integrated,tang2023solving}.
Nevertheless, this line of work is mostly restricted to  offline scenarios, i.e., tasks (and/or their release times) are assumed to be known.
In practice, for example in a sorting center, orders may come dynamically in real-time.

\begin{figure}[tb]
\centering
\includegraphics[width=\linewidth]{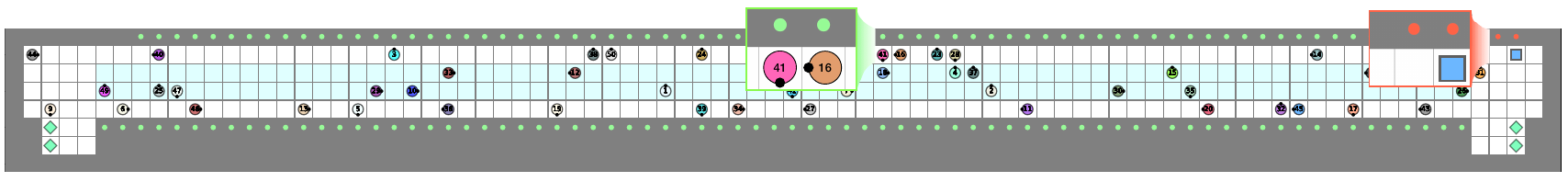}
\caption{A \textit{non-well-formed} instance 
currently deployed in Meituan warehouses. The white cells near \textsc{Green} dots are delivery ports, while the ones near \textsc{Red} dots are pickup ports. Colored circles heading to different directions with numbers are agents. The colored box (blue) is a pickup port currently assigned to the agent in the same color (ID 45 in the lower right area). Congestion happens a lot near the pickup ports.}
\label{fig:eg_non_wf}
\end{figure}

Besides, we draw attention to two seemingly minor but indeed fundamental aspects.
\textbf{For one}, the robots are usually abstracted to agents doing unit-cost unit-distance cardinal actions,
i.e., \{\texttt{stop}, \texttt{$\uparrow$}, \texttt{$\downarrow$}, \texttt{$\leftarrow$}, \texttt{$\rightarrow$}\}, what we term as the \texttt{Type}$\oplus$ robot model.
The planned paths are later post-processed to executable motions regarding kinematic constraints~\cite{honig2016multi} and action dependencies~\cite{honig2019persistent}, as a real-world robot has to rotate before going in a different direction.
Imaginably, when the rotational cost is not negligible compared to the translational cost, the quality of the plans computed for the \texttt{Type}$\oplus$ robot model will be largely compromised when instantiated to motions.
A candidate solution is to revisit and reimplement the existing algorithms over an alternative set of atomic actions \{\texttt{stop}, \texttt{forward}, \texttt{$\circlearrowright$90}, \texttt{$\circlearrowleft$90}\}, which we advocate in this paper as
the \texttt{Type}$\odot$ robot model~\cite{chan2024league, zhang2023efficient, jiang2024scaling}.
\textbf{For another}, most of the literature assumes the problem instance to be \textit{well-formed}~\cite{ma2017lifelong,liu2019task,xu2022multi} to guarantee completeness of their methods, which is actually a strong condition requiring that every agent can find a collision-free path to her current goal even if the others are stationary.
However, this assumption is often not met in modern warehouses. In particular, the instance (Figure~\ref{fig:eg_non_wf}) that we consider in this work does not satisfy this condition.









Considering the aforementioned issues,
we introduce a formal framework to study the combined problem that organically integrates task assignment and path finding in an online manner.
To solve the formalized problem,
we \textbf{first} develop lifelong path finding algorithms directly for the \texttt{Type}$\odot$ robot model (assuming an arbitrary task assigner), including those adapted from the existing literature and our new rule-based planner which performs both efficiently and effectively, even for non-well-formed instances.
\textbf{Secondly}, we propose a novel formulation that addresses the online problem of task assignment as optimally solving a Markov Decision Process~(MDP), with path planners as hyper-parameters.
Due to the complex state space and transition of the formulated MDP, we resort to approximated solutions by reinforcement learning (RL), as well as other non-trivial rule-based ones with insightful observations.
\textit{Simulation experiments on warehouse scenarios at Meituan, one of the largest shopping platforms in China, have shown that our system
(1)~takes only 83.77\% of the execution time needed for the currently deployed system at Meituan, outperforming other SOTA algorithms by 8.09\%; and
(2)~can achieve the same throughput with only 60\% of the agents of the current scale.}
We also draw an important lesson from this study
that both path finding and task assignment should fully exploit the warehouse layout, as it is normally fixed in a relatively long period of time after deployment, though the number of agents may still vary.
To this end, it might be more worthwhile to investigate layout-dependent-agent-independent solutions instead of entirely general-purpose solutions.

This paper is organized as follows.
We first list a few related areas in Section~\ref{sec:related}.
Then the problem formulation is provided in Section~\ref{sec:problem}.
We later present our system in two parts: the path planners in Section~\ref{sec:pf}, and the task assigners in Section~\ref{sec:ta}.
Experimental results are shown in Section~\ref{sec:exp}, mainly conducted for Meituan warehouse scenarios with various scales of agents.
We conclude the paper with a few insightful discussions in Section~\ref{sec:conclusion}.



\section{Related Work}
\label{sec:related}

\textbf{Path Finding.}
The study of MAPF aims to develop centralized planning algorithms.
In spite of the computational complexity being NP-hard in general~\cite{yu2013structure},
researchers have developed practically fast planners that can even solve instances with hundreds of agents within seconds.
Exemplars can be found via reduction to logic programs~\cite{erdem2013general},
prioritized planning~\cite{silver2005cooperative,ma2019searching,okumura2022priority},
conflict-based search~\cite{sharon2015conflict,li2021eecbs},
depth-first search~\cite{okumura2023lacam}, etc.
Most of them can be extended to the online version of the problem, i.e. MAPD,
where the goals assigned to agents are priorly unknown~\cite{ma2017lifelong,vsvancara2019online,li2021lifelong}.

\noindent
\textbf{Task Assignment.}
The earliest attempt is the formulation of \textit{Anonymous}-MAPF (AMAPF) that does not specify the exact goal that an agent must go to~\cite{stern2019multi}.
Compared with the labeled version, AMAPF can be solved in polynomial time,
via reduction to max-flow problems~\cite{yu2013multi}, or target swaps~\cite{okumura2023solving}.
As a generalization,
TAPF explicitly associates each agent with a team~\cite{ma2016optimal}, or with a set of candidate goals~\cite{honig2018conflict,tang2023solving},
and eventually computes a set of collision-free paths as well as the corresponding assignment matrix.
Another analogous formulation is called \textit{Multi-Goal} (MG-)MAPF~\cite{surynek2021multi,ijcai2024p0028} and its lifelong variant MG-MAPD~\cite{xu2022multi}, which also associates each agent with a set of goals, but the visiting order is pre-specified.



\textit{We also append some discussion on other less related areas to Appendix~\ref{apd:more_related_work}.}
Despite the rich literature, none of the above directly solves the online problem that a real-world automated warehouse is faced with, which well motivates this work.

\section{Problem Definition}
\label{sec:problem}


We consider a set of agents $N$, moving along a 4-neighbor grid map given as an undirected graph $G = (V, E)$,
where $V$ is the set of vertices and $E$ is the set of unit-cost edges.
Let $P, D\subseteq V$ denote the set of pickup ports and the set of delivery ports, respectively.
Usually, these two sets are disjoint and are specified alongside the map graph.

Let $k$ starting from 0 denote any arbitrary timestep (to be distinguished from the later notations of tasks).
At each timestep $k$, the local \textit{agent-state} of agent $i$, denoted as $\phi_i^k$, is a 3-tuple consisting of
her current location $l_i^k\in V$,
direction $d_i^k\in \{n,s,w,e\}$,
and goal $g_i^k \in P\cup D$.
$\Phi_i$ is the set of all possible local states of agent $i$, and consequently $\Phi = \prod_{i\in N} \Phi_i$ is the set of all possible \textit{joint agent-states}.
Each agent is associated with a set of four unit-cost actions $A=\{\texttt{stop}, \texttt{forward}, \texttt{$\circlearrowright$90}, \texttt{$\circlearrowleft$90}\}$ (called the \texttt{Type}$\odot$ robot model), with their usual meaning specified using the deterministic function $move$.
For example,
$move(((3,28), w), \texttt{$\circlearrowright$90}) \to ((3,28), n)$, and 
$move(((3,28), e), \texttt{forward}) \to ((3,29), e)$.
While planning paths for agents, we need to avoid the following two types of collisions~\cite{stern2019multi} for any pair of agents $(i, j)$: 1) vertex conflict, i.e., $l_i^k = l_j^k$; and 2) edge conflict, i.e., $l_i^k = l_j^{k+1} \land l_j^{k} = l_i^{k+1}$.



The so-called tasks are composed of a sequence $I$ of typed items. 
Each item $\iota \in I$ is associated with a type $t\in T$.
A back-end demand database specifies for each type a subset of delivery ports that items of the type need to be delivered to. 
Here we model such a database as a lookup table $L: T\mapsto 2^D$.
As $L$ will be changed in real-time, we also let $L^k$ denote the demand database at timestep~$k$.
When an agent has finished her last delivery job and returned to a pickup port at timestep $k$,
an item, say of type $t$, will be loaded onto this idle agent.
The system will then check the lookup table $L^k$, choose one target delivery port $g\in L^k(t)$ to assign to this agent, and delete this demand, i.e. $L^{k+1}(t) = L^k(t)\backslash \{g\}$.

Also, we have an assignment table $\eta$ that keeps track of which one of the loaded items is assigned to which agent, i.e., $\eta(\iota) = i$ means the item $\iota$ is currently carried by agent $i$.
An item $\iota$ is \textit{delivered} if there exists a timestep $k$ such that $l_{\eta(\iota)}^k = g_{\eta(\iota)}^k$, i.e., the agent carrying this item has reached her goal.
Upon successful delivery, the item $\iota$ will be deleted from the entry of $\eta$.
As $\eta$ is being changed  over time, we also use the time-indexed version $\eta^t$.
 

We assume (1) $L$ has recorded the demands of a long enough period of time, and therefore, will not be enlarged;
and (2) an item will be appended to the system
only when it is loaded onto an agent.



With the above notations, we formally define the dynamics of the whole system as a deterministic \textit{system-transition} function over \textit{system-states}.
\begin{enumerate}
	\item
	A \textit{system-state} $\psi$ is a tuple consisting of the joint agent-state $\phi = \{\phi_i\}_{i \in N}$, the lookup table $L$, and the assignment table $\eta$ at that moment. 
	Let $\Psi$ denote all possible system states.
	Among them, there is an initial system-state $\psi^0 = (\phi_1^0, \cdots, \phi_N^0, L^0, \eta^0)$.
	
	\item
	The space of \textit{system-actions} is $\Omega = (A\cup P\cup D)^N$. That is, a \textit{system-action} $\omega\in \Omega$ is an ordered tuple of the atomic actions of agents where any of them can be substituted by an assignment decision.
	A system-action $\omega$ is \textit{executable} under a legal system-state $\psi$ iff
	\[
	\forall i \in N.\ 
	 [l_i \in V\backslash (P\cup D) \land \omega_i \in A] \lor
	 [l_i \in P \land \omega_i \in D] \lor
	[l_i \in D \land \omega_i \in P]
	\]
	
	\item $\Gamma: \Psi \times \Omega \mapsto \Psi$ is the \textit{system-transition} function, which means (1) if no agents are at the pickup/delivery ports, then the system proceeds by deterministically moving agents by their reported actions, which will not change the goal component $g_i$ in each agent-state $\phi_i$; or (2) if any agent arrives at any pickup (resp. delivery) port, then the  system  needs to re-assign the agent the next delivery (resp. pickup) port, which will change the goal of that particular agent to the corresponding new location and temporarily force her to stay in-place, and change the demand table $L$ as well as the assignment table $\eta$ accordingly.
\end{enumerate}

An additional minor assumption is, even if two agents arrive at two different pickup (resp. delivery) ports simultaneously,
they will eventually get assigned certain new ports within the next one single timestep one by one in a random order. We do not care about the case where one is waiting for a new-delivery assignment while another is waiting for a new-pickup assignment.

In summary, a \textit{problem instance} is a following tuple $\langle N, G, P, D, A, I, L\rangle$,
and consequently a \textit{principle solution} is threefold:
\begin{enumerate}
	\item $\pi_N: \Phi\mapsto A^N$ is the routing policy that outputs the next action for each agent given their current states. It is unnecessary to compute the entire $\pi_N$ completely upfront, instead, execution can be interleaved with replanning.
	\item $\pi_{D}: \Psi \times 2^D \mapsto D$ is the delivery selection policy which assigns an agent a delivery port among the candidates according to $L(t)$ when she is at one of the pickup ports and given an item of type $t$.
	\item $\pi_{P}: \Psi \mapsto P$ is the pickup selection policy which assigns an agent a pickup port to return to when she has finished the delivery.
\end{enumerate}
Note that (1) both $\pi_D$ and $\pi_P$ assign one new goal at a time, as we assumed before;
(2) both $\pi_D$ and $\pi_P$ will change the goal of that particular agent to the corresponding port,
while $\pi_N$ will not;
(3) if an agent is at a pickup or delivery port, then her agent-action, even if specified by $\pi_N$, will be overwritten to $\texttt{stop}$ by the decision of $\pi_D$ or $\pi_P$.

\begin{definition}[Feasibility]
Given any system-state $\psi$ and the system-transition $\Gamma$, the application of a solution policy $(\pi_N, \pi_D, \pi_P)$ deterministically outputs a successor system-state $\psi'$. If there is no aforementioned collision between $\psi$ and $\psi'$, then $(\pi_N, \pi_D, \pi_P)$ is a feasible solution.
\end{definition}



We finally define \textit{makespan} as our evaluation metric.
\begin{definition}[Makespan]
	Given a problem instance with its initial system-state $\psi^0$, and a feasible solution $(\pi_N, \pi_D, \pi_P)$, an execution trajectory will be generated by the sequential applications of the solution policy $\{\psi^0, \psi^1, \cdots\}$. The makespan is the minimum timestep $k$ such that every item in I is delivered at~$\psi^k$. 
\end{definition}

However, in real-world warehouses, the pickup ports are usually concentrated in a restricted local area for operational convenience, e.g., in the top right corner of Figure~\ref{fig:eg_non_wf}. Therefore, $\pi_{P}$ is normally implemented for the purpose of balancing the loads over all pickup ports.
\textbf{In this work, we merely aim at a \textit{practical solution} consisting of only $(\pi_N, \pi_D)$, assuming $\pi_{P}$ is given and is not part of the desired solution.}



\begin{example}[System pipeline]
As shown in Figure~\ref{fig:eg_non_wf}, $\textsc{Robot}_1 \sim \textsc{Robot}_{50}$ initially rest in random locations after the last system execution. Once the system is launched, each robot moves towards the pickup ports, $\textsc{Red}_1$ and $\textsc{Red}_2$. When $\textsc{Robot}_{31}$ reaches $\textsc{Red}_2$, the human operator loads a dozen eggs onto it. The system checks the demand database and finds three orders for a dozen eggs, with delivery ports $\textsc{Green}_{69}$, $\textsc{Green}_{142}$, and $\textsc{Green}_{83}$. After consideration, the system decides to send $\textsc{Robot}_{31}$ to $\textsc{Green}_{83}$ this time, planning a path while avoiding potential collisions, with subsequent deliveries to the other two ports.
\end{example}

One may see potential connections between our problem and the standard formulation MAPD in the existing literature, \textit{we postpone some remarks elaborating on the differences to Appendix~\ref{apd:relate_to_mapd}, due to the page limit}.
We clarify that in the rest of the paper, by ``path finding'' we mean to compute $\pi_N$, and by ``task assignment'' we mean to compute $\pi_D$.

\section{Path Finding}

\label{sec:pf}

In this section, we first review several representative algorithms that can plan collision-free paths in a lifelong fashion.
However, they are not always effective for resolving collisions under the \texttt{Type}$\odot$ robot model for non-well-formed instances like Figure~\ref{fig:eg_non_wf}.
To this end, we propose a simple-yet-powerful rule-based planner
 which is capable of efficiently and robustly moving robots without collisions or deadlocks.


%

\subsection{Existing Lifelong Path Finding Algorithms}

\noindent
\textbf{Prioritized Planning.}
A straightforward way is to prioritize path finding for each agent by assigning them  distinct priorities, known as \textit{Cooperative A$^*$} (CA$^*$)~\cite{silver2005cooperative}, which can also be extended to lifelong situations.
In descending order of priority, the agents will plan their paths one by one.
Once an agent with a higher priority has found her path, those $(state, time)$ pairs along the path will be \textit{reserved} for this particular agent.
All subsequent agents with lower priorities will view those reservations as states that are unreachable at the corresponding timesteps, i.e. as spatio-temporal obstacles.
Therefore, each agent will need to conduct optimal search over the joint space of state and time.
Understandably,
there is a chance that an agent with a lower priority cannot compute any feasible path given the preceding path computed by a higher-priority, which makes the algorithm itself incomplete.
This situation is even worse under the \texttt{Type}$\odot$ robot model, as an agent often needs to rotate in-place before going to an adjacent vertex, which adds extra difficulty of avoiding collisions. \textit{An illustrative example is provided in Appendix~\ref{apd:pf_notgood_eg}}.

\noindent
\textbf{Rolling Horizon Collision Resolution.}
A more systematic approach for lifelong path finding is to \textit{window} the search process~\cite{silver2005cooperative}.
This idea is further developed by~\cite{li2021lifelong} as 
the \textit{Rolling Horizon Collision Resolution} (RHCR) framework.
The framework takes two use-specified parameters: (1)~the replanning frequency $h$ and (2)~the length of the collision resolution window $w\geq h$, ensuring that no collisions occur within the next $w$ timesteps.
The framework is general enough to encompass most MAPF algorithms.
An example is to
extend conflict-based search (CBS)~\cite{sharon2015conflict,li2021eecbs} to the lifelong version using this RHCR framework, where the high-level constraint tree is expanded only if there are still collisions within the first $w$ timesteps, resulting in a much smaller constraint tree.
However, under the \texttt{Type}$\odot$ robot model, neighboring agents often require more timesteps to resolve collisions, especially in crowded local areas.
\textit{An example is provided in Appendix~\ref{apd:pf_notgood_eg}.}

\subsection{Our Rule-Based Solution: Touring With Early Exit}


Instead of doing deliberate planning, we here present a simple-yet-effective rule-based planner, named \textit{Touring With Early Exit} (later denoted as \textbf{Touring} for short).
As summarized in Algorithm~\ref{alg:touring},
this planner consists of three main rules \textsc{Rule1-touring}, \textsc{Rule2-early-exit}, and \textsc{Rule3-safe}.
We will explain them one by one.

\begin{algorithm}
    \caption{Touring with early exit}
    \label{alg:touring}
    \textbf{Input}: $states = (\{l_i\}_{i\in N}, \{d_i\}_{i\in N}, \{g_i\}_{i\in N})$\\
    \textbf{Parameter}: for any arbitrary timestep $k$, omitted below\\
    \textbf{Output}: next joint-action $actions$
    \begin{algorithmic}[1] 
        \STATE $actions \gets \textsc{Rule1-touring}(states)$
        \IF{$\textsc{Exists-deadlock}(states)$}
        	\STATE $actions \gets \textsc{Rule3-safe}(states, actions)$
        	\STATE \textbf{return} actions
        \ENDIF
        \STATE $actions \gets \textsc{Rule2-early-exit}(states, actions)$
        \STATE $actions \gets \textsc{Rule3-safe}(states, actions)$
        \STATE \textbf{return} actions
    \end{algorithmic}
\end{algorithm}

\begin{figure}[tb]
\centering
\includegraphics[width=100mm]{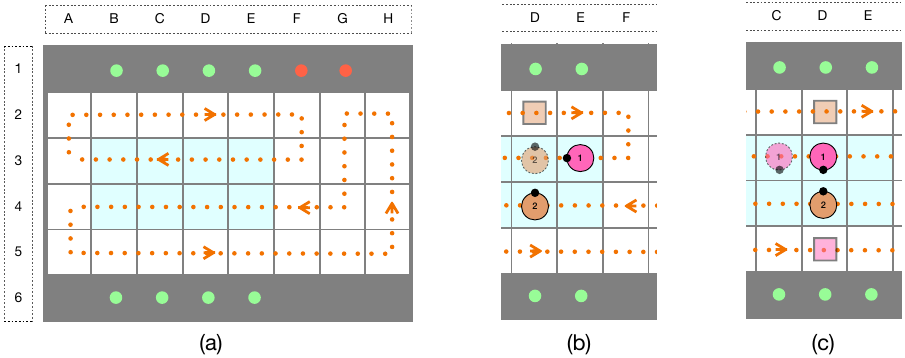}
\caption{\textsc{Rule1-Touring} (a) and two prioritized cases (b, c).
Colored boxes are the goals.}
\label{fig:touring}
\end{figure}

\textbf{Firstly}, for \textsc{Rule1-touring}, we partition the graph into disjoint \textit{Hamiltonian circuits} (HCs), termed as tours $\{\tau_p\}_{p\in P}$, ensuring that each tour covers distinct pickup ports.
Let $V_\tau\subset V$ denote the vertices in $\tau$, we have
$$\big( \forall p_1, p_2 \in P.\
p_1 \in \tau_1 \land p_2\in \tau_2 \land \tau_1 \neq \tau_2 \iff p_1 \neq p_2 \big) \land
\bigcup_{p\in P} V_{\tau_p} = V \land \bigcap_{p\in P} V_{\tau_p} = \emptyset $$
\textsc{Rule1-touring} further specifies the fixed direction along which agents will traverse the tour regardless of their goal locations, i.e. blind touring.
Figure~\ref{fig:touring}(a) shows an example with two tours (in dashed orange), one covering \textsc{F2} clockwise and the other covering \textsc{G2} counter-clockwise.
An agent may need more than one action at certain cells for touring, e.g., need a \texttt{$\circlearrowleft$90} followed by a \texttt{forward} at the corner~\textsc{A4}.

\textbf{Secondly}, for each tour $\tau$, $\textsc{Rule2-early-exit}$ marks certain vertices as turnings, where an agent currently in $\tau$ can exit the tour. The set of turnings is denoted as $V_\tau^{turn} \subseteq V_\tau$. 
An agent $i$ can \textit{exit} her tour $\tau$ if she is at the turning
that is the closest to her goal by choosing the next action of her shortest plan towards the goal,
or continue touring otherwise. Note that it may not be the case that $g_i \in V_\tau$, which may require agents to go across tours.
An exiting action is prioritized over a touring action.
Two agents who are exiting their own tours simultaneously but moving towards each other will be prioritized by their IDs:
the one with the larger ID will exit, while the other continues touring until reaching the next second-best turning.
The blue cells in Figure~\ref{fig:touring} represent the turnings of the respective tours, with \ref{fig:touring}(b) and \ref{fig:touring}(c) illustrating the two aforementioned prioritized cases.
These turnings can be either user-specified, or searched in terms of minimizing the makespan.
\textit{We provide an illustration of how the frequency of the turnings affects the eventual makespan in Appendix~\ref{apd:param_search}}.

\textbf{Finally}, \textsc{Rule3-safe}
is applied to revise those actions to collision-free ones.
For example, if a preceding agent is rotating, the following agent should not move \texttt{forward}; otherwise, collisions may occur.
Thus,
we design \textsc{Rule3-safe} conservatively: 
for each agent $i$ (1)~she observes her adjacent agents but assumes their actions specified by the prior rules may or may not be executed successfully, (2)~for either outcome, she checks whether her next action, if it is \texttt{forward}, will lead to a collision, (3) if any potential collision is detected, she revises her action to \texttt{stop}. 
Intuitively, this ensures that every agent maintains a safe distance from one another.
\textit{In fact, this conservative rule also prevents following-collisions}~\cite{stern2019multi}, which will not be discussed in this paper.

\textbf{Last but not least},
one may notice that
if there is a subset of agents forming a cycle where each one is about to go to the next location that is currently occupied by another agent in the cycle, \textsc{Rule3-safe} will overwrite the actions of all involved agents to \texttt{stop}, resulting in a deadlock.
Since the planner consisting of only the three main rules is merely a one-step reactive planner, the identical planning step will repeat indefinitely once a deadlock is formed.
Therefore, additional inspections need to be made (Line 2 in Algorithm~\ref{alg:touring}),
within which a depth-first search is conducted to see if any cycles (and thus the deadlock) exist.
 Once a deadlock is found, all the \textit{exiting} agents will be interrupted and resume \textit{touring}.
 By DFS for deadlock detection, we mean to construct a graph at each timestep, termed as $G_d = (V_d, E_d)$.
$V_d$ is the set of all agents, and $(i, j) \in E_d$ if $i, j$ are two adjacent agents and $move((l_i, d_i), \texttt{forward}) = (l_j, any)$.
If a cycle exists in $G_d$ (mostly caused by \textsc{Rule2-Early-Exit}), the by \textsc{Rule3-safe}, it is likely that every agent's next action will be overwritten to \texttt{stop}. In fact, \textbf{the existence of a cycle in $G_d$ is a necessary condition for the occurrence of a deadlock.} Once a cycle is detected, the whole system will switch to a safe touring-only mode by disabling \textsc{Rule2-Early-Exit} in the next step, thereby breaking the~deadlock.

Our \textbf{Touring} planner is sound as it eliminates potential collisions by implementing safety rules and avoiding deadlocks in real-time, but it might not be complete.
To ensure completeness, every item needs to be finally delivered to its destination: 1) an agent will eventually deliver her item if the delivery port is in the current tour, even in a touring-only mode; 2) if she has missed the current turning to go across tours, she has to seek another in the next few steps, or even the next touring round.
In principle, an agent may keep missing chances to go across tours, but it turns out to be a rare case.
In practice, an item is usually delivered in its first tour as we found in our experiments.

%
%

\subsection{Comparison for Path Finding Algorithms}
Before adding task assignment to the context, here we first conduct a brief comparison among the above path finding algorithms, assuming a sequence of goals arrives online.
We implement the lifelong CA$^*$ as a baseline for prioritized planning (denoted as \textbf{PP}),
and CBS under the RHCR framework with $h=1, w=5$ as a baseline for windowed search (denoted as \textbf{RHCR-CBS}).
We also implement two heuristics for the underlying single-agent search, namely $h_{slow}$, which merely computes the Manhattan distance between the current location and the goal, and $h_{fast}$, which additionally counts the minimum number of \texttt{$\circlearrowleft$90}/\texttt{$\circlearrowright$90} needed. Hence, here we have $2\times2=4$ combined baselines. \textit{We report the computing time, even for various scales, in Appendix~\ref{apd:comp_time}.} It turns out \textbf{RHCR-CBS-$h_{slow}$} is too costly for a multi-run evaluation.

As we mentioned, our testing environment (Figure~\ref{fig:eg_non_wf}) may not be well-formed.
\textbf{PP} may fail due to improper priority orderings.
\textbf{RHCR-CBS} may also take a long time for collision resolution, especially when there is a traffic jam near the pickup ports.
We treat a replanning of \textbf{RHCR-CBS} as failure if the number of leaf nodes in the high-level constraint tree exceeds 50, indicating severe congestion.
Once these two methods fail, they will be temporarily switched to \textbf{Touring}, and later be switched back.

\begin{figure}[tb]
\centering
\includegraphics[width=100mm]{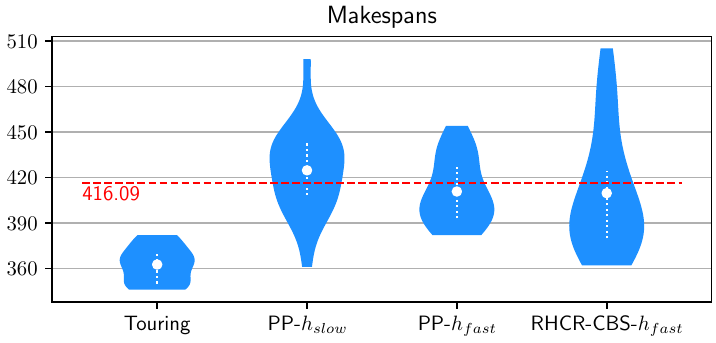}
\caption{The tested makespans of lifelong path finding algorithms in 50-agent Meituan warehouse scenarios. Dotted lines represent the 25-/75-quantiles, and white dots are the means. The means correspond to the leftmost column of the 50-agent scenario in Table~\ref{tab:eval_full}. \underline{416.09} is the reference makespan under Meituan's current system.}
\label{fig:pf_vp}
\end{figure}

In Figure~\ref{fig:pf_vp}, we present the entire distributions of the tested makespans over multiple runs.
As one can clearly see that our \textbf{Touring} planner largely outperforms the other three, and the computing time is entirely negligible compared to the others, as reported in Appendix~\ref{apd:comp_time}.
Besides, \textbf{RHCR-CBS} outperforms \textbf{PP} in most cases, though the average performances are close, as it poorly handles some extreme cases.


\section{Task Assignment}
\label{sec:ta}

In the offline setting where tasks are known \textit{a priori}
the assignment problem is well-studied~\cite{ma2016optimal,honig2018conflict,liu2019task,tang2023solving}, 
where the combinatorial search of the best task assignment is coupled with path finding.
However, when it comes to the online setting, it seems that the best approach so far is to greedily assign tasks at each decision point~\cite{ma2017lifelong,okumura2022priority}, i.e., to pick up the task so as to minimize the path costs from the current location to starting location of the task.
Projecting onto our settings, a greedy assignment is to select among those candidates the delivery port that is closest to the current location.
However, no evidence has witnessed that greedy assignments are rational and effective, given the fact that forthcoming tasks are totally unknown.
To this end, we extend this greedy strategy into a broader class of strategies, divided into three categories
(1)~stateless assignment, (2)~adaptive assignment, and (3)~predictive assignment.


\subsection{Stateless Assignment.}
As shown in Algorithm~\ref{alg:statless},
\textsc{MeasureFunc} is a user-specified function that encodes a measure between the location of the current agent waiting for assignment and the candidate delivery ports. Straightforward options are:
\begin{enumerate}
	\item \textbf{Shortest distances}. This reduces to the greedy strategies that select the closest delivery port.
	\item \textbf{Negative shortest distances}. This is equivalent to selecting the farthest delivery port. It is usually counter-intuitive, but makes some sense since it may alleviate congestion around the pickup ports, especially when the scale of the agents is large.
	\item \textbf{Random numbers}. It reduces to random assignments.
\end{enumerate}

\begin{algorithm}[!ht]
    \caption{Stateless assignment}
    \label{alg:statless}
    \textbf{Input}: agent $i$ waiting for assignment, item $\iota$ of type $t$, candidate delivery ports $L(t)$\\
    \textbf{Parameter}: any arbitrary timestep $k$ (omitted below)\\
    \textbf{Output}: A selected goal $\in L(t)$ 
    \begin{algorithmic}[1] 
        \STATE \textbf{return} $\arg\min_{g \in L(t)} \textsc{MeasureFunc}(g, l_i)$
    \end{algorithmic}
\end{algorithm}


\subsection{Adaptive Assignment.}
\begin{figure}
\centering
\includegraphics[width=110mm]{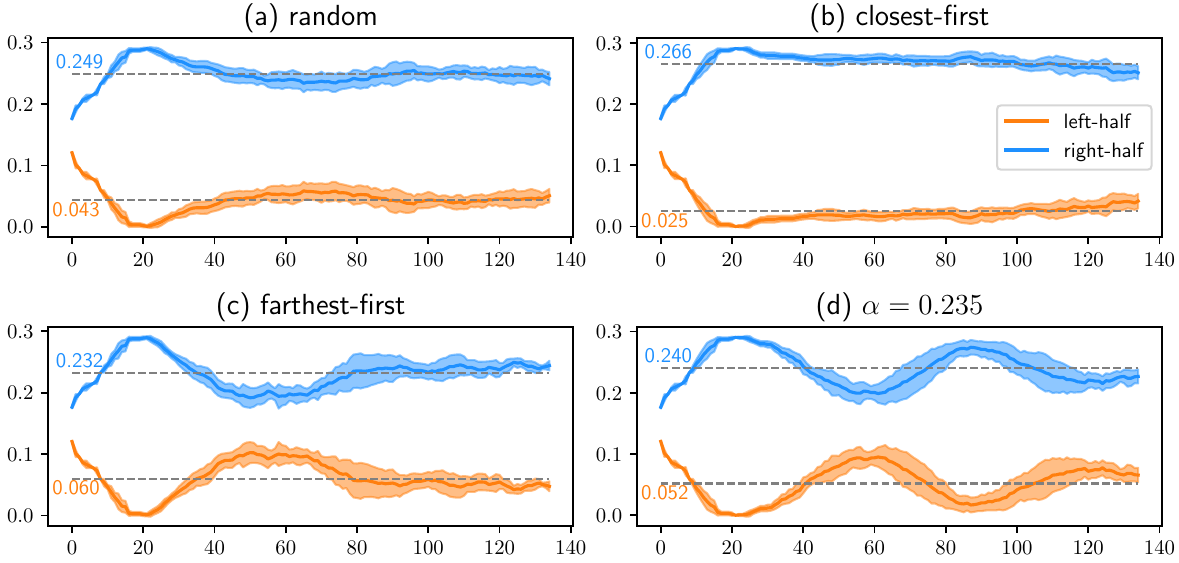}
\caption{The dynamics of the occupation ratios for different stateless and adaptive assignment strategies in 50-agent cases. Dashed lines represent the means.}
\vspace{-1mm}
\label{fig:occu_ratio}
\end{figure}

Stateless assignments do not make the most use of system-state information, e.g., the current locations of all agents.
As revealed in Figure~\ref{fig:occu_ratio}, the occupation ratio, defined as the fraction of the number of agents over the number of passable cells,
of the left half differs significantly from that of the right half.
The \textit{closest-first} strategy will inevitably cause high-level congestion around the pickup ports, while \textit{farthest-first} strategy unnecessarily sends agents to farther locations, even though it alleviates traffic jams.
The random strategy somehow
balances between the former two.

Inspired by this insight, Algorithm~\ref{alg:adaptive} further develops an adaptive version, which takes in a congestion threshold $\alpha$ and makes dynamic assignment decisions based on the current state. If there is a heavy traffic in the right half of the map, the system will send agents to farther goals, and similarly otherwise. One can clearly see in Figure~\ref{fig:occu_ratio}(d) that the occupation ratio fluctuates more responsively.

The threshold parameter $\alpha$ can be specified by users or searched in terms of minimizing the makespan.
\textit{We report comprehensive search results in
Figure~\ref{fig:alpha_bp} in Appendix~\ref{apd:param_search}.}

\begin{algorithm}
    \caption{Adaptive assignment}
    \label{alg:adaptive}
    \textbf{Input}: agent $i$ waiting for assignment, item $\iota$ of type $t$, candidate delivery ports $L(t)$, all agents' locations $\{l_i\}_{i\in N}$\\
    \textbf{Parameter}: A threshold $\alpha$, any timestep $k$ (omitted below)\\
    \textbf{Output}:  A selected goal $\in L(t)$ 
    \begin{algorithmic}[1] 
    	\STATE $occu_{l}, occu_{r} \gets \textsc{OccupationRatio}(\{l_i\}_{i\in N})$
    	\IF{$occu_{r} \leq \alpha$}
    		\STATE \textbf{return} $\arg\min_{g \in L(t)} \textsc{ShortestDistance}(g, l_i)$
		\ELSE
			\STATE \textbf{return} $\arg\max_{g \in L(t)} \textsc{ShortestDistance}(g, l_i)$
    	\ENDIF
    \end{algorithmic}
\end{algorithm}


\subsection{Predictive Assignment.}
\label{sec:ta_rl} 
One can further argue that purely reactive assignments like the above ones do not capture the dynamics of the system.
To this end, one needs to make good use of the underlying path finding module which may hint about the dynamic flow of the agent swarm.
A systematic way is to formulate the assignment problem as a Markov Decision Process (MDP), taking the path finding module, i.e. the routing policy $\pi_N$, as a hyperparameter.
The MDP is defined as a 5-tuple
$\langle\mathcal{S}, \mathcal{A}, T, R, \gamma\rangle$:
\begin{enumerate}
	\item \textbf{States} $\mathcal{S}$: each $S \in \mathcal{S} \subset \Psi$ is a collection of all system-states where there exists an agent at a pickup port waiting for a new-delivery assignment. We call these \textit{assignment states} to avoid ambiguity.
	\item \textbf{Actions} $\mathcal{A} = D$: all possible delivery ports. Given a loaded item of type $t$, the available actions are the delivery ports in $L(t)$.
	\item \textbf{Transition function} $T: \mathcal{S}\times \mathcal{A} \mapsto \Delta(\mathcal{S})$. Once a new delivery port is assigned to an agent, the system will proceed according to $\pi_N$ (and the given $\pi_P$) until the next assignment state. 
	\item \textbf{Reward function} $R: \mathcal{S}\times \mathcal{A} \times \mathcal{S} \mapsto \mathbb{R}$. The reward is the negative time cost spent between two assignment states. Note that the reward signals are quite ``delayed'', in the sense that for two adjacent assignment states, the immediate reward received at the latter one might not reflect the delivery cost for the task assigned in the former state (it is usually not delivered yet). Nevertheless, the total accumulated rewards of an episode is indeed the negative makespan to complete the attended item sequence.
	\item \textbf{Discount factor} $\gamma \in (0, 1)$.
\end{enumerate}
Note that to solve the above MDP is to search for the policy $\pi_D$ while fixing $\pi_N$.
By definition, once the optimal policy $\pi_D^*$ is found, it will instruct the system to assign tasks at any possible assignment state, and therefore, the initial resting locations of agents
do not matter.
We adopt PPO~\cite{schulman2017proximal} as the RL algorithm to solve the above MDP, \textit{and will postpone some training details and remarks to Appendix~\ref{apd:rl_train}}. 


\section{Experimental Results}
\label{sec:exp}


\begin{table}[ht]
\caption{Evaluation results. The numbers are the average makespans ($\downarrow$).
The reference makespan is \underline{416.09} by the currently deployed system at Meituan. 
For each path planner, the result performed by the best task assigner is marked in \textbf{bold}. 
The scenario in \textcolor{teal}{teal} is the current scale at Meituan. 
Those in \textcolor{orange}{orange} represent the best combinations at each scale.
Some cells are marked in \textcolor{gray}{gray} as the RL policies are not explicitly trained for those scenarios.
For the adaptive assignment strategies, we report the top-3 ones along with the corresponding thresholds in superscripts.
For the RL strategies, besides the best ones of average performances, we also present the best ones of best-case (resp. worst-case) performances, along with the corresponding best-case (resp. worst-case) makespans in superscripts.}
\centering
\resizebox{1\textwidth}{!}{
\begin{tabular}{@{}lrrrrrrrrr@{}}
\toprule
(30 agents)                  & \textbf{Random}      & \textbf{Closest}      & \textbf{Farthest}     & \textbf{Adapt$^{1st}$} & \textbf{Adapt$^{2nd}$} & \textbf{Adapt$^{3rd}$} & \textbf{RL$^{avg}$}  & \textbf{RL$^{best}$}  & \textbf{RL$^{worst}$}  \\ \midrule
\textbf{Touring (ours)}      & 467.00               & \textcolor{orange}{$\textbf{412.90}$}               & 530.85               & $443.45^{0.158}$       & $454.15^{0.152}$       & $454.15^{0.155}$       & 425.00                  & $425.00^{412}$          & $425.00^{439}$          \\
\textbf{PP $h_{slow}$}       & 659.45               & $\textbf{550.75}$               & 678.50               & $587.70^{0.158}$        & $590.10^{0.149}$        & $605.30^{0.146}$        & \cellcolor[HTML]{EFEFEF}569.45               & \cellcolor[HTML]{EFEFEF}$569.45^{482}$       & \cellcolor[HTML]{EFEFEF}$569.45^{762}$       \\
\textbf{PP $h_{fast}$}       & 641.30               & $\textbf{561.50}$               & 681.50               & $589.10^{0.152}$        & $589.50^{0.155}$        & $610.20^{0.149}$        & \cellcolor[HTML]{EFEFEF}588.40                & \cellcolor[HTML]{EFEFEF}$588.40^{492}$        & \cellcolor[HTML]{EFEFEF}$588.40^{800}$        \\
\textbf{RHCR-CBS $h_{fast}$} & 645.00               & $\textbf{539.75}$               & 726.05               & $645.20^{0.152}$        & $646.20^{0.155}$        & $655.10^{0.146}$        & \cellcolor[HTML]{EFEFEF}641.95               & \cellcolor[HTML]{EFEFEF}$641.95^{495}$       & \cellcolor[HTML]{EFEFEF}$641.95^{800}$     \vspace{1.5mm}  \\
(40 agents)                  &                      &                      &                      &                        &                        &                        &                      &                      &                      \\ \midrule
\textbf{Touring (ours)}      & 392.10               & 376.30               & 422.70               & $382.40^{0.219}$        & $387.30^{0.211}$        & $387.30^{0.215}$        & \textcolor{orange}{$\textbf{372.05}$}               & $372.05^{348}$       & $383.65^{399}$       \\
\textbf{PP $h_{slow}$}       & 474.50               & $\textbf{427.10}$               & 518.35               & $443.50^{0.215}$        & $447.20^{0.211}$        & $449.05^{0.207}$       & \cellcolor[HTML]{EFEFEF}452.80                & \cellcolor[HTML]{EFEFEF}$452.80^{425}$        & \cellcolor[HTML]{EFEFEF}$473.90^{565}$        \\
\textbf{PP $h_{fast}$}       & 467.00               & $\textbf{426.70}$               & 516.40               & $443.70^{0.219}$        & $449.15^{0.215}$       & $451.25^{0.211}$       & \cellcolor[HTML]{EFEFEF}445.20                & \cellcolor[HTML]{EFEFEF}$445.20^{417}$        & \cellcolor[HTML]{EFEFEF}$475.45^{535}$       \\
\textbf{RHCR-CBS $h_{fast}$} & 463.00               & 444.95               & 523.75               & $438.85^{0.211}$       & $438.85^{0.215}$       & $443.10^{0.219}$        & \cellcolor[HTML]{EFEFEF}$\textbf{431.55}$               & \cellcolor[HTML]{EFEFEF}$431.55^{394}$       & \cellcolor[HTML]{EFEFEF}$447.05^{481}$    \vspace{1.5mm}   \\
\textcolor{teal}{\textbf{(50 agents)}}         &                      &                      &                      &                        &                        &                        &                      &                      &                      \\ \midrule
\textcolor{teal}{\textbf{Touring (ours)}}      & 362.55               & 358.35               & 375.15               & \textcolor{orange}{$\textbf{348.55}^{0.235}$}       & $349.35^{0.265}$       & $349.80^{0.240}$         & 350.15               & $352.70^{316}$        & $350.15^{363}$       \\
\textcolor{teal}{\textbf{PP $h_{slow}$}}       & 424.65               & 392.85               & 435.45               & $\textbf{388.35}^{0.280}$        & $392.65^{0.275}$       & $400.55^{0.255}$       & \cellcolor[HTML]{EFEFEF}409.95               & \cellcolor[HTML]{EFEFEF}$397.10^{359}$        & \cellcolor[HTML]{EFEFEF}$409.95^{509}$       \\
\textcolor{teal}{\textbf{PP $h_{fast}$}}       & 410.70               & $\textbf{390.15}$               & 434.20               & $396.70^{0.280}$         & $399.40^{0.265}$        & $400.15^{0.260}$        & \cellcolor[HTML]{EFEFEF}398.25               & \cellcolor[HTML]{EFEFEF}$402.70^{361}$        & \cellcolor[HTML]{EFEFEF}$398.25^{444}$       \\
\textcolor{teal}{\textbf{RHCR-CBS $h_{fast}$}} & 409.60               & 401.00               & 415.90               & $384.25^{0.280}$        & $385.20^{0.275}$        & $386.10^{0.265}$        & \cellcolor[HTML]{EFEFEF}384.90                & \cellcolor[HTML]{EFEFEF}$\textbf{382.20}^{363}$        & \cellcolor[HTML]{EFEFEF}$384.90^{468}$    \vspace{1.5mm}    \\
(60 agents)                  & \multicolumn{1}{l}{} & \multicolumn{1}{l}{} & \multicolumn{1}{l}{} & \multicolumn{1}{l}{}   & \multicolumn{1}{l}{}   & \multicolumn{1}{l}{}   & \multicolumn{1}{l}{} & \multicolumn{1}{l}{} & \multicolumn{1}{l}{} \\ \midrule
\textbf{Touring (ours)}      & 350.60                & 352.50                & 352.40                & \textcolor{orange}{$\textbf{335.50}^{0.281}$}        & $337.10^{0.293}$        & $337.10^{0.299}$        & 342.70               & $342.70^{308}$        & $342.70^{359}$        \\
\textbf{PP $h_{slow}$}       & 390.90                & 380.45               & 411.00                & $\textbf{369.35}^{0.287}$       & $370.20^{0.293}$        & $373.10^{0.329}$        & \cellcolor[HTML]{EFEFEF}375.10                & \cellcolor[HTML]{EFEFEF}$375.10^{345}$        & \cellcolor[HTML]{EFEFEF}$375.10^{403}$        \\
\textbf{PP $h_{fast}$}       & 394.15               & 382.80                & 397.15               & $\textbf{371.75}^{0.299}$       & $372.65^{0.329}$       & $378.45^{0.311}$       & \cellcolor[HTML]{EFEFEF}391.05               & \cellcolor[HTML]{EFEFEF}$391.05^{356}$       & \cellcolor[HTML]{EFEFEF}$391.05^{499}$       \\
\textbf{RHCR-CBS $h_{fast}$} & 372.50                & 370.00                & 375.90                & $\textbf{357.35}^{0.305}$       & $360.55^{0.287}$       & $360.85^{0.323}$       & \cellcolor[HTML]{EFEFEF}372.85               & \cellcolor[HTML]{EFEFEF}$372.85^{354}$       & \cellcolor[HTML]{EFEFEF}$372.85^{469}$   \vspace{1.5mm}    \\
(70 agents)                  &                      &                      &                      &                        &                        &                        &                      &                      &                      \\ \midrule
\textbf{Touring (ours)}      & 346.45               & 354.65               & 344.50                & \textcolor{orange}{$\textbf{333.40}^{0.353}$}        & $333.60^{0.339}$        & $334.10^{0.360}$         & 338.80                & $338.80^{308}$        & $338.80^{354}$        \\
\textbf{PP $h_{slow}$}       & 375.95               & 381.15               & 393.85               & $374.95^{0.325}$       & $375.40^{0.388}$        & $375.95^{0.304}$       & \cellcolor[HTML]{EFEFEF}$\textbf{373.50}$                & \cellcolor[HTML]{EFEFEF}$373.50^{347}$        & \cellcolor[HTML]{EFEFEF}$373.50^{394}$        \\
\textbf{PP $h_{fast}$}       & 371.25               & 372.10                & 372.10                & $\textbf{364.95}^{0.332}$       & $364.95^{0.360}$        & $365.35^{0.304}$       & \cellcolor[HTML]{EFEFEF}390.90                & \cellcolor[HTML]{EFEFEF}$390.90^{340}$        & \cellcolor[HTML]{EFEFEF}$390.90^{513}$        \\
\textbf{RHCR-CBS $h_{fast}$} & 362.50                & 377.20                & 365.55               & $\textbf{351.90}^{0.367}$        & $353.30^{0.353}$        & $354.10^{0.381}$        & \cellcolor[HTML]{EFEFEF}362.20                & \cellcolor[HTML]{EFEFEF}$362.20^{337}$        & \cellcolor[HTML]{EFEFEF}$362.20^{435}$        \\ \bottomrule
\end{tabular}
}
\label{tab:eval_full}
\end{table}

In this section, we report the main experimental results in Table~\ref{tab:eval_full}, conducted on Meituan simulated warehouse (Figure~\ref{fig:eg_non_wf}).
The online sequences of items are retrieved from roughly 5-minute segments of the system's log containing around $140$ items 
of approximately 50 types, 
while the demand database $L$ is made from mobile orders made by the customers in a longer period of around 6 hours. 
The two dimensions of Table~\ref{tab:eval_full} represent the choices of path planners and task assigners, respectively.
The numbers are the average \textit{makespans} over multiple runs  with agents initialized in random locations.
In each run, the system is required to deliver a sequence of items of a fixed length that arrives online.
In other words, we evaluate the average cost it takes to accomplish the same amount of throughput, for each pair of path planners and task assigners.
For the adaptive assignment strategies, we report the top-3 ones along with the corresponding thresholds in superscripts.
For the RL strategies, besides the best ones for average performances, we also present the best ones for best-case (resp. worst-case) performances, along with the corresponding best-case (resp. worst-case) makespans in superscripts.
We only train RL policies over the \textbf{Touring} planner, since the others are not fast enough and will take a tremendous amount of time for RL training.
However, we can slightly abuse a trained assignment policy by testing it with the other three path planners as the state spaces are same.

\underline{As an overview},
our \textbf{Touring} planner outperforms the other three regardless of the task assigner.
As for the task assigner,
(1)~when the number of agents is $\geq 50$,
adaptive strategies are surprisingly effective, even slightly better than RL ones, and the \textit{closest-first} strategy is not necessarily better than the \textit{farthest-first} strategy.
(2)~when the number of agents is $< 50$,
it might be redundant to use adaptive strategies as the density of agents is quite low; instead, stateless ones or RL ones are better choices.
Although RL strategies can achieve comparable performances in practice (even optimal performance in theory if trained well),
it depends on the user whether the training cost is a worthwhile effort.


\underline{Looking closer}, we point out two insights:
\begin{enumerate}
\item \textit{Time efficiency}.
Regarding the current scale of Meituan (50 agents),
our system only needs 83.77\% of the makespan to deliver the same amount of throughput, compared to the company's current system $^{(348.55 / 416.09)}$, outperforming the best$^{(382.20 / 416.09)}$ among the rest by 8.09\%.

\item \textit{Economic efficiency}.
Note that there is a continuing improvement$^{(348.55 \rightarrow 335.50)}$ while increasing the number of agents to 60. However, the marginal gain of further increasing to 70 agents is negligible$^{(335.50 \rightarrow 333.40)}$.
In fact, only 30 agents can fulfill the current throughput with even slightly shorter time$^{(412.90)}$, resulting in a 40\% reduction in fixed costs for purchasing robots. 
\end{enumerate}

We additionally point out that, while the \textit{closer-first} strategy seems effective in the experiments and is widely adopted in the literature, such fixed-priority methods are not desirable in practice.
Similar to processor scheduling in OS, a significant drawback is that certain tasks may continually be preempted, i.e., a delivery port gets no item for a long operational period.



%
%


\section{Conclusion and Future Work}
\label{sec:conclusion}

In this paper, we conduct a case study on the real-world problem of warehouse automation by combining lifelong multi-robot path finding and dynamic task assignment in an online fashion.
As a result, we manage to speed up package delivery given the current scale at Meituan,
and also identify potential profitable upgrades of the system.
An important lesson from this study is that given the layout of the warehouse, once deployed, is normally fixed in a relatively long period of time, it is worthwhile to have both the routing module and the assignment module that take advantage of the layout.
However, both modules should be general enough to account for the varying number of robots available.
We conclude the paper by clarifying some concerns for generalizability and pointing out some future work.

\textbf{Generalizability and Limitation.}
\textit{For the routing module}, in this study, we identify two disjoint HCs that cover all delivery ports, with each circuit servicing one pickup port. Additional rules are implemented to prioritize agents crossing between the tours. While a similar strategy can be applied in general, determining the optimal approach may be nontrivial, and thus an intriguing topic for future research. For instance, one needs to consider the trade-off between using multiple HCs to cover different pickup ports versus employing one single large HC that covers all delivery and pickup ports.
\textit{For the assignment module},
1)~for stateless assignments, one can replace the distance measure with another suitable one; 2) for adaptive assignments, one can divide the whole map into a few subareas to monitor congestion (we presented the case of two monitored subareas), and 3) for predictive assignments, the MDP formulation is layout-independent, solely taking in a given pathfinder.
A limitation of this work is that we search an assignment policy w.r.t. a fixed routing policy, which is an open-loop control. The next step is to couple the search of these two, though it will be computationally challenging.

\section*{Acknowledgments}
This work was supported in part by 
Meituan Academy of Robotics Shenzhen (MARS).
We thank our colleagues from Meituan\textemdash Yuan Liu, Shumin Zhang, Fuqiang Hei, and Yuxuan Zhang\textemdash for their invaluable assistance and collaboration in this project, and colleagues from HKUST\textemdash Yuxin Pan, Yangfan Wu and Xiaomeng Zhu\textemdash for their insightful feedback during the revision of this paper.
We also appreciate the constructive comments from the anonymous reviewers.

\bibliographystyle{eptcs}
\bibliography{generic}

\begin{thebibliography}{10}
\providecommand{\bibitemdeclare}[2]{}
\providecommand{\surnamestart}{}
\providecommand{\surnameend}{}
\providecommand{\urlprefix}{Available at }
\providecommand{\url}[1]{\texttt{#1}}
\providecommand{\href}[2]{\texttt{#2}}
\providecommand{\urlalt}[2]{\href{#1}{#2}}
\providecommand{\doi}[1]{doi:\urlalt{https://doi.org/#1}{#1}}
\providecommand{\eprint}[1]{arXiv:\urlalt{https://arxiv.org/abs/#1}{#1}}
\providecommand{\bibinfo}[2]{#2}

\bibitemdeclare{article}{braekers2016vehicle}
\bibitem{braekers2016vehicle}
\bibinfo{author}{Kris \surnamestart Braekers\surnameend},
  \bibinfo{author}{Katrien \surnamestart Ramaekers\surnameend} \&
  \bibinfo{author}{Inneke \surnamestart Van~Nieuwenhuyse\surnameend}
  (\bibinfo{year}{2016}): \emph{\bibinfo{title}{The vehicle routing problem:
  State of the art classification and review}}.
\newblock {\slshape \bibinfo{journal}{Computers \& industrial engineering}}
  \bibinfo{volume}{99}, pp. \bibinfo{pages}{300--313}.
\newblock \urlprefix\url{https://doi.org/10.1016/j.cie.2015.12.007}.

\bibitemdeclare{inproceedings}{chan2024league}
\bibitem{chan2024league}
\bibinfo{author}{Shao-Hung \surnamestart Chan\surnameend}, \bibinfo{author}{Zhe
  \surnamestart Chen\surnameend}, \bibinfo{author}{Teng \surnamestart
  Guo\surnameend}, \bibinfo{author}{Han \surnamestart Zhang\surnameend},
  \bibinfo{author}{Yue \surnamestart Zhang\surnameend}, \bibinfo{author}{Daniel
  \surnamestart Harabor\surnameend}, \bibinfo{author}{Sven \surnamestart
  Koenig\surnameend}, \bibinfo{author}{Cathy \surnamestart Wu\surnameend} \&
  \bibinfo{author}{Jingjin \surnamestart Yu\surnameend} (\bibinfo{year}{2024}):
  \emph{\bibinfo{title}{The League of Robot Runners Competition: Goals,
  Designs, and Implementation}}.
\newblock In: {\slshape \bibinfo{booktitle}{ICAPS 2024 System's Demonstration
  track}}.

\bibitemdeclare{article}{chen2021integrated}
\bibitem{chen2021integrated}
\bibinfo{author}{Zhe \surnamestart Chen\surnameend}, \bibinfo{author}{Javier
  \surnamestart Alonso-Mora\surnameend}, \bibinfo{author}{Xiaoshan
  \surnamestart Bai\surnameend}, \bibinfo{author}{Daniel~D \surnamestart
  Harabor\surnameend} \& \bibinfo{author}{Peter~J \surnamestart
  Stuckey\surnameend} (\bibinfo{year}{2021}): \emph{\bibinfo{title}{Integrated
  task assignment and path planning for capacitated multi-agent pickup and
  delivery}}.
\newblock {\slshape \bibinfo{journal}{IEEE Robotics and Automation Letters}}
  \bibinfo{volume}{6}(\bibinfo{number}{3}), pp. \bibinfo{pages}{5816--5823},
  \doi{10.1109/LRA.2021.3074883}.

\bibitemdeclare{article}{damani2021primal}
\bibitem{damani2021primal}
\bibinfo{author}{Mehul \surnamestart Damani\surnameend},
  \bibinfo{author}{Zhiyao \surnamestart Luo\surnameend},
  \bibinfo{author}{Emerson \surnamestart Wenzel\surnameend} \&
  \bibinfo{author}{Guillaume \surnamestart Sartoretti\surnameend}
  (\bibinfo{year}{2021}): \emph{\bibinfo{title}{PRIMAL$_2$: Pathfinding via
  reinforcement and imitation multi-agent learning-lifelong}}.
\newblock {\slshape \bibinfo{journal}{IEEE Robotics and Automation Letters}}
  \bibinfo{volume}{6}(\bibinfo{number}{2}), pp. \bibinfo{pages}{2666--2673},
  \doi{10.1109/LRA.2021.3062803}.

\bibitemdeclare{inproceedings}{erdem2013general}
\bibitem{erdem2013general}
\bibinfo{author}{Esra \surnamestart Erdem\surnameend}, \bibinfo{author}{Doga
  \surnamestart Kisa\surnameend}, \bibinfo{author}{Umut \surnamestart
  Oztok\surnameend} \& \bibinfo{author}{Peter \surnamestart
  Sch{\"u}ller\surnameend} (\bibinfo{year}{2013}): \emph{\bibinfo{title}{A
  general formal framework for pathfinding problems with multiple agents}}.
\newblock In: {\slshape \bibinfo{booktitle}{Proceedings of the AAAI Conference
  on Artificial Intelligence}}, \bibinfo{volume}{27}, pp.
  \bibinfo{pages}{290--296}.
\newblock \urlprefix\url{https://doi.org/10.1609/aaai.v27i1.8592}.

\bibitemdeclare{article}{ginsberg1989universal}
\bibitem{ginsberg1989universal}
\bibinfo{author}{Matthew~L \surnamestart Ginsberg\surnameend}
  (\bibinfo{year}{1989}): \emph{\bibinfo{title}{Universal planning: An (almost)
  universally bad idea}}.
\newblock {\slshape \bibinfo{journal}{AI magazine}}
  \bibinfo{volume}{10}(\bibinfo{number}{4}), pp. \bibinfo{pages}{40--40}.
\newblock \urlprefix\url{https://doi.org/10.1609/aimag.v10i4.964}.

\bibitemdeclare{inproceedings}{honig2018conflict}
\bibitem{honig2018conflict}
\bibinfo{author}{Wolfgang \surnamestart H{\"o}nig\surnameend},
  \bibinfo{author}{Scott \surnamestart Kiesel\surnameend},
  \bibinfo{author}{Andrew \surnamestart Tinka\surnameend},
  \bibinfo{author}{Joseph~W \surnamestart Durham\surnameend} \&
  \bibinfo{author}{Nora \surnamestart Ayanian\surnameend}
  (\bibinfo{year}{2018}): \emph{\bibinfo{title}{Conflict-Based Search with
  Optimal Task Assignment}}.
\newblock In: {\slshape \bibinfo{booktitle}{Proceedings of the 17th
  International Conference on Autonomous Agents and MultiAgent Systems}}, pp.
  \bibinfo{pages}{757--765}.
\newblock \urlprefix\url{https://dl.acm.org/doi/abs/10.5555/3237383.3237495}.

\bibitemdeclare{article}{honig2019persistent}
\bibitem{honig2019persistent}
\bibinfo{author}{Wolfgang \surnamestart H{\"o}nig\surnameend},
  \bibinfo{author}{Scott \surnamestart Kiesel\surnameend},
  \bibinfo{author}{Andrew \surnamestart Tinka\surnameend},
  \bibinfo{author}{Joseph~W \surnamestart Durham\surnameend} \&
  \bibinfo{author}{Nora \surnamestart Ayanian\surnameend}
  (\bibinfo{year}{2019}): \emph{\bibinfo{title}{Persistent and robust execution
  of MAPF schedules in warehouses}}.
\newblock {\slshape \bibinfo{journal}{IEEE Robotics and Automation Letters}}
  \bibinfo{volume}{4}(\bibinfo{number}{2}), pp. \bibinfo{pages}{1125--1131},
  \doi{10.1109/LRA.2019.2894217}.

\bibitemdeclare{inproceedings}{honig2016multi}
\bibitem{honig2016multi}
\bibinfo{author}{Wolfgang \surnamestart H{\"o}nig\surnameend},
  \bibinfo{author}{TK~\surnamestart Kumar\surnameend}, \bibinfo{author}{Liron
  \surnamestart Cohen\surnameend}, \bibinfo{author}{Hang \surnamestart
  Ma\surnameend}, \bibinfo{author}{Hong \surnamestart Xu\surnameend},
  \bibinfo{author}{Nora \surnamestart Ayanian\surnameend} \&
  \bibinfo{author}{Sven \surnamestart Koenig\surnameend}
  (\bibinfo{year}{2016}): \emph{\bibinfo{title}{Multi-agent path finding with
  kinematic constraints}}.
\newblock In: {\slshape \bibinfo{booktitle}{Proceedings of the International
  Conference on Automated Planning and Scheduling}}, \bibinfo{volume}{26}, pp.
  \bibinfo{pages}{477--485}.
\newblock \urlprefix\url{https://doi.org/10.1609/icaps.v26i1.13796}.

\bibitemdeclare{inproceedings}{huang2022closer}
\bibitem{huang2022closer}
\bibinfo{author}{Shengyi \surnamestart Huang\surnameend} \&
  \bibinfo{author}{Santiago \surnamestart Onta{\~n}{\'o}n\surnameend}
  (\bibinfo{year}{2022}): \emph{\bibinfo{title}{A Closer Look at Invalid Action
  Masking in Policy Gradient Algorithms}}.
\newblock In: {\slshape \bibinfo{booktitle}{The International FLAIRS Conference
  Proceedings}}, \bibinfo{volume}{35}.
\newblock \urlprefix\url{https://doi.org/10.32473/flairs.v35i.130584}.

\bibitemdeclare{article}{jain1999deterministic}
\bibitem{jain1999deterministic}
\bibinfo{author}{Anant~Singh \surnamestart Jain\surnameend} \&
  \bibinfo{author}{Sheik \surnamestart Meeran\surnameend}
  (\bibinfo{year}{1999}): \emph{\bibinfo{title}{Deterministic job-shop
  scheduling: Past, present and future}}.
\newblock {\slshape \bibinfo{journal}{European journal of operational
  research}} \bibinfo{volume}{113}(\bibinfo{number}{2}), pp.
  \bibinfo{pages}{390--434}.
\newblock \urlprefix\url{https://doi.org/10.1016/S0377-2217(98)00113-1}.

\bibitemdeclare{inproceedings}{jiang2024scaling}
\bibitem{jiang2024scaling}
\bibinfo{author}{He~\surnamestart Jiang\surnameend}, \bibinfo{author}{Yulun
  \surnamestart Zhang\surnameend}, \bibinfo{author}{Rishi \surnamestart
  Veerapaneni\surnameend} \& \bibinfo{author}{Jiaoyang \surnamestart
  Li\surnameend} (\bibinfo{year}{2024}): \emph{\bibinfo{title}{Scaling lifelong
  multi-agent path finding to more realistic settings: Research challenges and
  opportunities}}.
\newblock In: {\slshape \bibinfo{booktitle}{Proceedings of the International
  Symposium on Combinatorial Search}}, \bibinfo{volume}{17}, pp.
  \bibinfo{pages}{234--242}.
\newblock \urlprefix\url{https://doi.org/10.1609/socs.v17i1.31565}.

\bibitemdeclare{inproceedings}{li2021eecbs}
\bibitem{li2021eecbs}
\bibinfo{author}{Jiaoyang \surnamestart Li\surnameend},
  \bibinfo{author}{Wheeler \surnamestart Ruml\surnameend} \&
  \bibinfo{author}{Sven \surnamestart Koenig\surnameend}
  (\bibinfo{year}{2021}): \emph{\bibinfo{title}{Eecbs: A bounded-suboptimal
  search for multi-agent path finding}}.
\newblock In: {\slshape \bibinfo{booktitle}{Proceedings of the AAAI conference
  on artificial intelligence}}, \bibinfo{volume}{35}, pp.
  \bibinfo{pages}{12353--12362}.
\newblock \urlprefix\url{https://doi.org/10.1609/aaai.v35i14.17466}.

\bibitemdeclare{inproceedings}{li2021lifelong}
\bibitem{li2021lifelong}
\bibinfo{author}{Jiaoyang \surnamestart Li\surnameend}, \bibinfo{author}{Andrew
  \surnamestart Tinka\surnameend}, \bibinfo{author}{Scott \surnamestart
  Kiesel\surnameend}, \bibinfo{author}{Joseph~W \surnamestart
  Durham\surnameend}, \bibinfo{author}{TK~Satish \surnamestart
  Kumar\surnameend} \& \bibinfo{author}{Sven \surnamestart Koenig\surnameend}
  (\bibinfo{year}{2021}): \emph{\bibinfo{title}{Lifelong multi-agent path
  finding in large-scale warehouses}}.
\newblock In: {\slshape \bibinfo{booktitle}{Proceedings of the AAAI Conference
  on Artificial Intelligence}}, \bibinfo{volume}{35}, pp.
  \bibinfo{pages}{11272--11281}.
\newblock \urlprefix\url{https://doi.org/10.1609/aaai.v35i13.17344}.

\bibitemdeclare{inproceedings}{liu2019task}
\bibitem{liu2019task}
\bibinfo{author}{Minghua \surnamestart Liu\surnameend}, \bibinfo{author}{Hang
  \surnamestart Ma\surnameend}, \bibinfo{author}{Jiaoyang \surnamestart
  Li\surnameend} \& \bibinfo{author}{Sven \surnamestart Koenig\surnameend}
  (\bibinfo{year}{2019}): \emph{\bibinfo{title}{Task and path planning for
  multi-agent pickup and delivery}}.
\newblock In: {\slshape \bibinfo{booktitle}{Proceedings of the International
  Joint Conference on Autonomous Agents and Multiagent Systems (AAMAS)}}.
\newblock \urlprefix\url{https://dl.acm.org/doi/abs/10.5555/3306127.3331816}.

\bibitemdeclare{inproceedings}{ma2019searching}
\bibitem{ma2019searching}
\bibinfo{author}{Hang \surnamestart Ma\surnameend}, \bibinfo{author}{Daniel
  \surnamestart Harabor\surnameend}, \bibinfo{author}{Peter~J \surnamestart
  Stuckey\surnameend}, \bibinfo{author}{Jiaoyang \surnamestart Li\surnameend}
  \& \bibinfo{author}{Sven \surnamestart Koenig\surnameend}
  (\bibinfo{year}{2019}): \emph{\bibinfo{title}{Searching with consistent
  prioritization for multi-agent path finding}}.
\newblock In: {\slshape \bibinfo{booktitle}{Proceedings of the AAAI conference
  on artificial intelligence}}, \bibinfo{volume}{33}, pp.
  \bibinfo{pages}{7643--7650}.
\newblock \urlprefix\url{https://doi.org/10.1609/aaai.v33i01.33017643}.

\bibitemdeclare{inproceedings}{ma2016optimal}
\bibitem{ma2016optimal}
\bibinfo{author}{Hang \surnamestart Ma\surnameend} \& \bibinfo{author}{Sven
  \surnamestart Koenig\surnameend} (\bibinfo{year}{2016}):
  \emph{\bibinfo{title}{Optimal Target Assignment and Path Finding for Teams of
  Agents}}.
\newblock In: {\slshape \bibinfo{booktitle}{Proceedings of the 2016
  International Conference on Autonomous Agents \& Multiagent Systems}}, pp.
  \bibinfo{pages}{1144--1152}.
\newblock \urlprefix\url{https://dl.acm.org/doi/abs/10.5555/2936924.2937092}.

\bibitemdeclare{inproceedings}{ma2017lifelong}
\bibitem{ma2017lifelong}
\bibinfo{author}{Hang \surnamestart Ma\surnameend}, \bibinfo{author}{Jiaoyang
  \surnamestart Li\surnameend}, \bibinfo{author}{TK~Satish \surnamestart
  Kumar\surnameend} \& \bibinfo{author}{Sven \surnamestart Koenig\surnameend}
  (\bibinfo{year}{2017}): \emph{\bibinfo{title}{Lifelong Multi-Agent Path
  Finding for Online Pickup and Delivery Tasks}}.
\newblock In: {\slshape \bibinfo{booktitle}{Proceedings of the 16th Conference
  on Autonomous Agents and MultiAgent Systems}}, pp. \bibinfo{pages}{837--845}.
\newblock \urlprefix\url{https://dl.acm.org/doi/10.5555/3091125.3091243}.

\bibitemdeclare{article}{manne1960job}
\bibitem{manne1960job}
\bibinfo{author}{Alan~S \surnamestart Manne\surnameend} (\bibinfo{year}{1960}):
  \emph{\bibinfo{title}{On the job-shop scheduling problem}}.
\newblock {\slshape \bibinfo{journal}{Operations research}}
  \bibinfo{volume}{8}(\bibinfo{number}{2}), pp. \bibinfo{pages}{219--223}.
\newblock \urlprefix\url{https://doi.org/10.1287/opre.8.2.219}.

\bibitemdeclare{inproceedings}{okumura2023lacam}
\bibitem{okumura2023lacam}
\bibinfo{author}{Keisuke \surnamestart Okumura\surnameend}
  (\bibinfo{year}{2023}): \emph{\bibinfo{title}{Lacam: Search-based algorithm
  for quick multi-agent pathfinding}}.
\newblock In: {\slshape \bibinfo{booktitle}{Proceedings of the AAAI Conference
  on Artificial Intelligence}}, \bibinfo{volume}{37}, pp.
  \bibinfo{pages}{11655--11662}.
\newblock \urlprefix\url{https://doi.org/10.1609/aaai.v37i10.26377}.

\bibitemdeclare{article}{okumura2023solving}
\bibitem{okumura2023solving}
\bibinfo{author}{Keisuke \surnamestart Okumura\surnameend} \&
  \bibinfo{author}{Xavier \surnamestart D{\'e}fago\surnameend}
  (\bibinfo{year}{2023}): \emph{\bibinfo{title}{Solving simultaneous target
  assignment and path planning efficiently with time-independent execution}}.
\newblock {\slshape \bibinfo{journal}{Artificial Intelligence}}
  \bibinfo{volume}{321}, p. \bibinfo{pages}{103946}.
\newblock \urlprefix\url{https://doi.org/10.1016/j.artint.2023.103946}.

\bibitemdeclare{article}{okumura2022priority}
\bibitem{okumura2022priority}
\bibinfo{author}{Keisuke \surnamestart Okumura\surnameend},
  \bibinfo{author}{Manao \surnamestart Machida\surnameend},
  \bibinfo{author}{Xavier \surnamestart D{\'e}fago\surnameend} \&
  \bibinfo{author}{Yasumasa \surnamestart Tamura\surnameend}
  (\bibinfo{year}{2022}): \emph{\bibinfo{title}{Priority inheritance with
  backtracking for iterative multi-agent path finding}}.
\newblock {\slshape \bibinfo{journal}{Artificial Intelligence}}
  \bibinfo{volume}{310}, p. \bibinfo{pages}{103752}.
\newblock \urlprefix\url{https://doi.org/10.1016/j.artint.2022.103752}.

\bibitemdeclare{article}{sartoretti2019primal}
\bibitem{sartoretti2019primal}
\bibinfo{author}{Guillaume \surnamestart Sartoretti\surnameend},
  \bibinfo{author}{Justin \surnamestart Kerr\surnameend},
  \bibinfo{author}{Yunfei \surnamestart Shi\surnameend}, \bibinfo{author}{Glenn
  \surnamestart Wagner\surnameend}, \bibinfo{author}{TK~Satish \surnamestart
  Kumar\surnameend}, \bibinfo{author}{Sven \surnamestart Koenig\surnameend} \&
  \bibinfo{author}{Howie \surnamestart Choset\surnameend}
  (\bibinfo{year}{2019}): \emph{\bibinfo{title}{Primal: Pathfinding via
  reinforcement and imitation multi-agent learning}}.
\newblock {\slshape \bibinfo{journal}{IEEE Robotics and Automation Letters}}
  \bibinfo{volume}{4}(\bibinfo{number}{3}), pp. \bibinfo{pages}{2378--2385},
  \doi{10.1109/LRA.2019.2903261}.

\bibitemdeclare{inproceedings}{schoppers1987universal}
\bibitem{schoppers1987universal}
\bibinfo{author}{Marcel \surnamestart Schoppers\surnameend}
  (\bibinfo{year}{1987}): \emph{\bibinfo{title}{Universal Plans for Reactive
  Robots in Unpredictable Environments.}}
\newblock In: {\slshape \bibinfo{booktitle}{IJCAI}}, \bibinfo{volume}{87},
  \bibinfo{organization}{Citeseer}, pp. \bibinfo{pages}{1039--1046}.
\newblock \urlprefix\url{https://dl.acm.org/doi/abs/10.5555/1625995.1626091}.

\bibitemdeclare{article}{schulman2017proximal}
\bibitem{schulman2017proximal}
\bibinfo{author}{John \surnamestart Schulman\surnameend},
  \bibinfo{author}{Filip \surnamestart Wolski\surnameend},
  \bibinfo{author}{Prafulla \surnamestart Dhariwal\surnameend},
  \bibinfo{author}{Alec \surnamestart Radford\surnameend} \&
  \bibinfo{author}{Oleg \surnamestart Klimov\surnameend}
  (\bibinfo{year}{2017}): \emph{\bibinfo{title}{Proximal policy optimization
  algorithms}}.
\newblock {\slshape \bibinfo{journal}{arXiv preprint arXiv:1707.06347}}.

\bibitemdeclare{article}{sharon2015conflict}
\bibitem{sharon2015conflict}
\bibinfo{author}{Guni \surnamestart Sharon\surnameend}, \bibinfo{author}{Roni
  \surnamestart Stern\surnameend}, \bibinfo{author}{Ariel \surnamestart
  Felner\surnameend} \& \bibinfo{author}{Nathan~R \surnamestart
  Sturtevant\surnameend} (\bibinfo{year}{2015}):
  \emph{\bibinfo{title}{Conflict-based search for optimal multi-agent
  pathfinding}}.
\newblock {\slshape \bibinfo{journal}{Artificial intelligence}}
  \bibinfo{volume}{219}, pp. \bibinfo{pages}{40--66}.
\newblock \urlprefix\url{https://doi.org/10.1016/j.artint.2014.11.006}.

\bibitemdeclare{inproceedings}{silver2005cooperative}
\bibitem{silver2005cooperative}
\bibinfo{author}{David \surnamestart Silver\surnameend} (\bibinfo{year}{2005}):
  \emph{\bibinfo{title}{Cooperative pathfinding}}.
\newblock In: {\slshape \bibinfo{booktitle}{Proceedings of the aaai conference
  on artificial intelligence and interactive digital entertainment}},
  \bibinfo{volume}{1}, pp. \bibinfo{pages}{117--122}.
\newblock \urlprefix\url{https://doi.org/10.1609/aiide.v1i1.18726}.

\bibitemdeclare{inproceedings}{stern2019multi}
\bibitem{stern2019multi}
\bibinfo{author}{Roni \surnamestart Stern\surnameend}, \bibinfo{author}{Nathan
  \surnamestart Sturtevant\surnameend}, \bibinfo{author}{Ariel \surnamestart
  Felner\surnameend}, \bibinfo{author}{Sven \surnamestart Koenig\surnameend},
  \bibinfo{author}{Hang \surnamestart Ma\surnameend}, \bibinfo{author}{Thayne
  \surnamestart Walker\surnameend}, \bibinfo{author}{Jiaoyang \surnamestart
  Li\surnameend}, \bibinfo{author}{Dor \surnamestart Atzmon\surnameend},
  \bibinfo{author}{Liron \surnamestart Cohen\surnameend},
  \bibinfo{author}{TK~\surnamestart Kumar\surnameend} et~al.
  (\bibinfo{year}{2019}): \emph{\bibinfo{title}{Multi-agent pathfinding:
  Definitions, variants, and benchmarks}}.
\newblock In: {\slshape \bibinfo{booktitle}{Proceedings of the International
  Symposium on Combinatorial Search}}, \bibinfo{volume}{10}, pp.
  \bibinfo{pages}{151--158}.
\newblock \urlprefix\url{https://doi.org/10.1609/socs.v10i1.18510}.

\bibitemdeclare{inproceedings}{surynek2021multi}
\bibitem{surynek2021multi}
\bibinfo{author}{Pavel \surnamestart Surynek\surnameend}
  (\bibinfo{year}{2021}): \emph{\bibinfo{title}{Multi-goal multi-agent path
  finding via decoupled and integrated goal vertex ordering}}.
\newblock In: {\slshape \bibinfo{booktitle}{Proceedings of the AAAI Conference
  on Artificial Intelligence}}, \bibinfo{volume}{35}, pp.
  \bibinfo{pages}{12409--12417}.
\newblock \urlprefix\url{https://doi.org/10.1609/aaai.v35i14.17472}.

\bibitemdeclare{inproceedings}{vsvancara2019online}
\bibitem{vsvancara2019online}
\bibinfo{author}{Ji{\v{r}}{\'\i} \surnamestart {\v{S}}vancara\surnameend},
  \bibinfo{author}{Marek \surnamestart Vlk\surnameend}, \bibinfo{author}{Roni
  \surnamestart Stern\surnameend}, \bibinfo{author}{Dor \surnamestart
  Atzmon\surnameend} \& \bibinfo{author}{Roman \surnamestart
  Bart{\'a}k\surnameend} (\bibinfo{year}{2019}): \emph{\bibinfo{title}{Online
  multi-agent pathfinding}}.
\newblock In: {\slshape \bibinfo{booktitle}{Proceedings of the AAAI conference
  on artificial intelligence}}, \bibinfo{volume}{33}, pp.
  \bibinfo{pages}{7732--7739}.
\newblock \urlprefix\url{https://doi.org/10.1609/aaai.v33i01.33017732}.

\bibitemdeclare{inproceedings}{ijcai2024p0028}
\bibitem{ijcai2024p0028}
\bibinfo{author}{Mingkai \surnamestart Tang\surnameend},
  \bibinfo{author}{Yuanhang \surnamestart Li\surnameend},
  \bibinfo{author}{Hongji \surnamestart Liu\surnameend},
  \bibinfo{author}{Yingbing \surnamestart Chen\surnameend},
  \bibinfo{author}{Ming \surnamestart Liu\surnameend} \& \bibinfo{author}{Lujia
  \surnamestart Wang\surnameend} (\bibinfo{year}{2024}):
  \emph{\bibinfo{title}{MGCBS: An Optimal and Efficient Algorithm for Solving
  Multi-Goal Multi-Agent Path Finding Problem}}.
\newblock In \bibinfo{editor}{Kate \surnamestart Larson\surnameend}, editor:
  {\slshape \bibinfo{booktitle}{Proceedings of the Thirty-Third International
  Joint Conference on Artificial Intelligence, {IJCAI-24}}},
  \bibinfo{publisher}{International Joint Conferences on Artificial
  Intelligence Organization}, pp. \bibinfo{pages}{249--256}.
\newblock \urlprefix\url{https://doi.org/10.24963/ijcai.2024/28}.
\newblock \bibinfo{note}{Main Track}.

\bibitemdeclare{inproceedings}{tang2023solving}
\bibitem{tang2023solving}
\bibinfo{author}{Yimin \surnamestart Tang\surnameend},
  \bibinfo{author}{Zhongqiang \surnamestart Ren\surnameend},
  \bibinfo{author}{Jiaoyang \surnamestart Li\surnameend} \&
  \bibinfo{author}{Katia \surnamestart Sycara\surnameend}
  (\bibinfo{year}{2023}): \emph{\bibinfo{title}{Solving multi-agent target
  assignment and path finding with a single constraint tree}}.
\newblock In: {\slshape \bibinfo{booktitle}{2023 International Symposium on
  Multi-Robot and Multi-Agent Systems (MRS)}}, \bibinfo{organization}{IEEE},
  pp. \bibinfo{pages}{8--14}, \doi{10.1109/MRS60187.2023.10416794}.

\bibitemdeclare{book}{toth2014vehicle}
\bibitem{toth2014vehicle}
\bibinfo{author}{Paolo \surnamestart Toth\surnameend} \&
  \bibinfo{author}{Daniele \surnamestart Vigo\surnameend}
  (\bibinfo{year}{2014}): \emph{\bibinfo{title}{Vehicle routing: problems,
  methods, and applications}}.
\newblock \bibinfo{publisher}{SIAM}.
\newblock \urlprefix\url{https://doi.org/10.1137/1.9781611973594.fm}.

\bibitemdeclare{inproceedings}{xu2022multi}
\bibitem{xu2022multi}
\bibinfo{author}{Qinghong \surnamestart Xu\surnameend},
  \bibinfo{author}{Jiaoyang \surnamestart Li\surnameend}, \bibinfo{author}{Sven
  \surnamestart Koenig\surnameend} \& \bibinfo{author}{Hang \surnamestart
  Ma\surnameend} (\bibinfo{year}{2022}): \emph{\bibinfo{title}{Multi-goal
  multi-agent pickup and delivery}}.
\newblock In: {\slshape \bibinfo{booktitle}{2022 IEEE/RSJ International
  Conference on Intelligent Robots and Systems (IROS)}},
  \bibinfo{organization}{IEEE}, pp. \bibinfo{pages}{9964--9971},
  \doi{10.1109/IROS47612.2022.9981785}.

\bibitemdeclare{inproceedings}{yu2013structure}
\bibitem{yu2013structure}
\bibinfo{author}{Jingjin \surnamestart Yu\surnameend} \&
  \bibinfo{author}{Steven \surnamestart LaValle\surnameend}
  (\bibinfo{year}{2013}): \emph{\bibinfo{title}{Structure and intractability of
  optimal multi-robot path planning on graphs}}.
\newblock In: {\slshape \bibinfo{booktitle}{Proceedings of the AAAI Conference
  on Artificial Intelligence}}, \bibinfo{volume}{27}, pp.
  \bibinfo{pages}{1443--1449}.
\newblock \urlprefix\url{https://doi.org/10.1609/aaai.v27i1.8541}.

\bibitemdeclare{inproceedings}{yu2013multi}
\bibitem{yu2013multi}
\bibinfo{author}{Jingjin \surnamestart Yu\surnameend} \&
  \bibinfo{author}{Steven~M \surnamestart LaValle\surnameend}
  (\bibinfo{year}{2013}): \emph{\bibinfo{title}{Multi-agent path planning and
  network flow}}.
\newblock In: {\slshape \bibinfo{booktitle}{Algorithmic Foundations of Robotics
  X: Proceedings of the Tenth Workshop on the Algorithmic Foundations of
  Robotics}}, \bibinfo{organization}{Springer}, pp. \bibinfo{pages}{157--173}.
\newblock \urlprefix\url{https://doi.org/10.1007/978-3-642-36279-8_10}.

\bibitemdeclare{inproceedings}{zhang2023efficient}
\bibitem{zhang2023efficient}
\bibinfo{author}{Yue \surnamestart Zhang\surnameend}, \bibinfo{author}{Daniel
  \surnamestart Harabor\surnameend}, \bibinfo{author}{Pierre \surnamestart
  Le~Bodic\surnameend} \& \bibinfo{author}{Peter~J \surnamestart
  Stuckey\surnameend} (\bibinfo{year}{2023}): \emph{\bibinfo{title}{Efficient
  Multi Agent Path Finding with Turn Actions}}.
\newblock In: {\slshape \bibinfo{booktitle}{Proceedings of the International
  Symposium on Combinatorial Search}}, \bibinfo{volume}{16}, pp.
  \bibinfo{pages}{119--127}.
\newblock \urlprefix\url{https://doi.org/10.1609/socs.v16i1.27290}.

\bibitemdeclare{inproceedings}{zhu2023computing}
\bibitem{zhu2023computing}
\bibinfo{author}{Fengming \surnamestart Zhu\surnameend} \&
  \bibinfo{author}{Fangzhen \surnamestart Lin\surnameend}
  (\bibinfo{year}{2025}): \emph{\bibinfo{title}{Computing Universal Plans for
  Partially Observable Multi-Agent Routing Using Answer Set Programming}}.
\newblock In: {\slshape \bibinfo{booktitle}{Proceedings of the International
  Conference on Logic Programming (ICLP)}}, \bibinfo{volume}{to appear}.
\newblock \urlprefix\url{https://arxiv.org/abs/2305.16203}.

\end{thebibliography}

\onecolumn
\appendix

\section{Relation to MAPD}
\label{apd:relate_to_mapd}

In MAPD, an online task $t_i$ is characterized by a pickup port $s_i$ and a delivery port $g_i$ with a priorly unknown release time. Once an agent becomes idle, she will select one task $t^* = (s^*, g^*)$ of her best interest from the released ones, and then plan a path from her current location to $g^*$ through $s^*$.
Mapping to our settings, an agent becomes idle only when she arrives at a pickup port, and shall then be assigned one delivery port from the candidates, say $\{g_1, \cdots, g_k\}$.
Suppose the system will simply pair each delivery port with a pickup port immediately, for which the particular agent will return to after the delivery. Then it is equivalent to, in the language of MAPD, releasing $k$ tasks $\{(g_1, \pi_P(g_1)), \cdots, (g_k, \pi_P(g_k)\}$.
However, after choosing one from the $k$ tasks and assigning it to an agent, the rest $(k-1)$ tasks will be temporarily removed, or ``deactivated'', from the pool of released tasks until the next item of the same type arrives.

\section{More Discussion on Related Work}
\label{apd:more_related_work}

The problem presented in this paper pertains to some other important areas in planning and operations research.

\textbf{Universal Planning.}
Unlike the idealized one-shot MAPF, fully automating real-world warehouses requires lifelong path finding.
However, most of the existing work~\cite{ma2017lifelong,vsvancara2019online,li2021lifelong,xu2022multi} still focuses on the solution concept as a set of collision-free paths, which is a sequence of joint actions.
Such a solution concept is vulnerable if there is any uncertainty, e.g. unknown future goals, or even system contingencies.
We argue that one can turn to the solution concept of universal plans~\cite{schoppers1987universal,ginsberg1989universal}.
Although universal plans are even harder to compute, there are some exemplars using multi-agent reinforcement learning~\cite{sartoretti2019primal,damani2021primal}, or via reduction to logic programs~\cite{zhu2023computing}.

\textbf{Scheduling.}
One may also notice the analogy between TAPF and job-shop scheduling problems (JSSP)~\cite{manne1960job,jain1999deterministic} or vehicle routing problems (VRP)~\cite{toth2014vehicle,braekers2016vehicle}.
However, there are at least two key differences: (1) job durations in JSSP and route lengths in VRP are usually known in advance and (2) the execution of jobs or routes is independent of each other.
Neither of these two conditions holds in TAPF, especially when the tasks are released online.

\section{Side Effects of the \texttt{Type}$\odot$ Robot Model}
\label{apd:pf_notgood_eg}

\begin{figure*}[!ht]
    \begin{subfigure}[b]{\textwidth}
        \includegraphics[width=95mm]{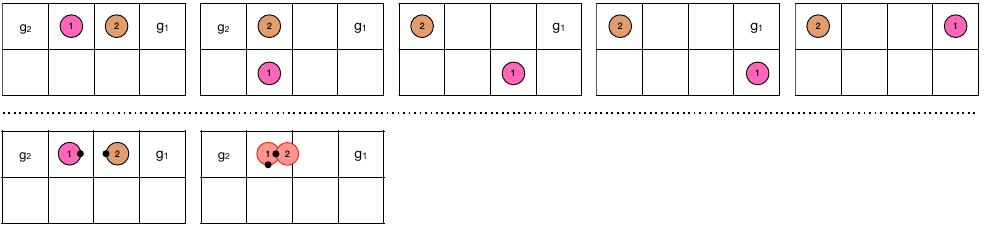} 
        \caption{Cooperative A$^*$ may fail for the \texttt{Type}$\odot$ robot model}
    \end{subfigure}
    \vspace{0.5mm}
    
    \begin{subfigure}[b]{\textwidth}
        \includegraphics[width=152mm]{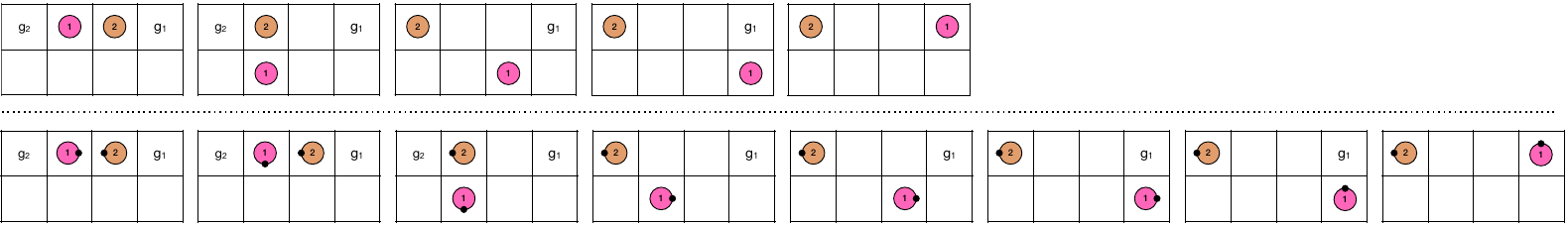}
        \caption{Conflict-based search works but takes more timesteps (same consequence by priority-based search in this particular case).}
    \end{subfigure}
    \vspace{0.5mm}
    
    \begin{subfigure}[b]{\textwidth}
        \includegraphics[width=114mm]{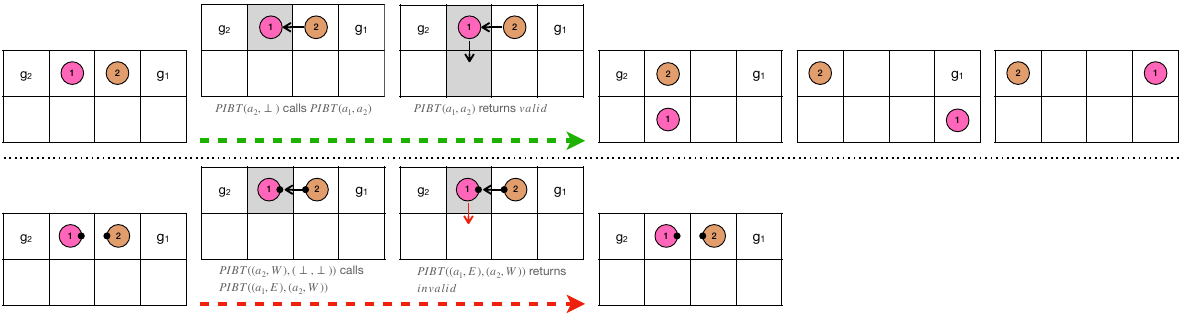}
        \caption{PIBT fails for the \texttt{Type}$\odot$ robot model.}
    \end{subfigure}

    \caption{Examples showing the added difficulty of resolving collisions with the \texttt{Type}$\odot$ robot model.}
    \label{fig:pf_notgood_eg}
\end{figure*}

When the state of an agent is lifted from pure locations to (location, direction) pairs, there will be extra difficulty resolving collisions. We here show three examples for (a)~\textit{cooperative A$^*$} (CA$^*$)~\cite{silver2005cooperative}, (b)~\textit{conflict-based search} (CBS)~\cite{sharon2015conflict}, and (c)~\textit{priority inheritance with backtracking} (PIBT)~\cite{okumura2022priority}, respectively.
\begin{enumerate}
	\item Figure~\ref{fig:pf_notgood_eg}(a) shows a case where CA$^*$ fails for the \texttt{Type}$\odot$ robot model. Suppose agent 2 is prioritized over agent 1, then agent 1 will move away immediately under the \texttt{Type}$\oplus$ robot model. However, under the \texttt{Type}$\odot$ robot model, agent 1 has to rotate first and thus cannot manage to avoid collision at the very next timestep.
	\item Figure~\ref{fig:pf_notgood_eg}(b) shows a case where it takes CBS a longer time  to resolve collisions under the \texttt{Type}$\odot$ robot model. The main reason is still due to the rotational cost. Similarly for the execution of \textit{priority-based search} (PBS)~\cite{ma2019searching}. 
	\item Figure~\ref{fig:pf_notgood_eg}(c) shows a failed case due to deeper theoretical reasons.
	Instead of performing any best-first search, PIBT repeats one-timestep planning until the terminal state, and therefore, it needs a crucial lemma to make sure the total number of execution steps is always bounded (see \textbf{Lemma~1} in~\cite{okumura2022priority}), i.e., \textit{at each timestep the agent with the highest priority will manage to move one step closer to her goal}.
Nevertheless, when the states of each agent are lifted from only locations to (location, direction) pairs, this lemma no longer holds as a counter-example is provided in Figure~\ref{fig:pf_notgood_eg}(c).
\end{enumerate}

%
%

\section{Computing Time}
\label{apd:comp_time}

We report the planning time per step in Table~\ref{tab:plan_time}.
Experiments are conducted on a
MacBook Air with Apple M2 CPU and 16 GB memory.
The planners are all implemented in Python, therefore, those numbers are merely for relative comparisons within this work.

\begin{table}[!ht]
\small
\centering
\begin{tabular}{@{}lrrrrr@{}}
\toprule
                             & \textbf{30} & \textbf{40} & \textbf{50} & \textbf{60} & \textbf{70} \\ \midrule
\textbf{Touring (ours)}      & $\textbf{0.008}$       & $\textbf{0.012}$       & $\textbf{0.018}$       & $\textbf{0.025}$       & $\textbf{0.032}$       \\
\textbf{PP $h_{fast}$}       & 0.066       & 0.115       & 0.225       & 0.500       & 0.713       \\
\textbf{PP $h_{slow}$}       & 0.169       & 0.287       & 0.448       & 0.892       & 1.028       \\
\textbf{RHCR-CBS $h_{fast}$} & 0.765       & 0.718       & 1.842       & 2.642       & 2.448       \\
\textbf{RHCR-CBS $h_{slow}$} & 3.023       & 3.070       & 7.081       & 7.821       & 8.952       \\ \bottomrule
\end{tabular}
\caption{Planning time per step in seconds, implemented in Python.}
\label{tab:plan_time}
\end{table}

\section{Parameter Search}
\label{apd:param_search}
\begin{figure*}[!h]
\centering
\includegraphics[width=\linewidth]{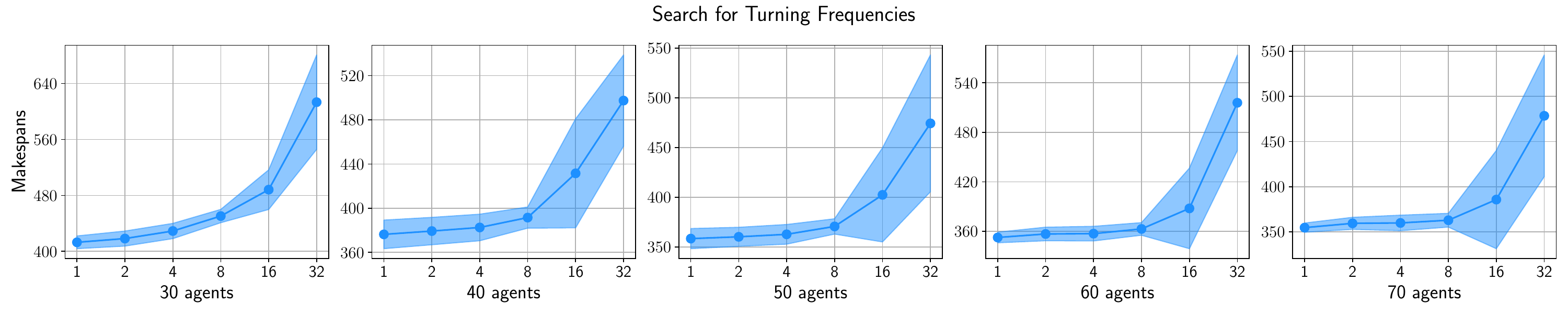}
\caption{Makespans over different turning frequency in various scales of agents in Meituan warehouse simulation. The X-axis means ``there will be a turning every $x$ cells''.}
\label{fig:turning_freq}
\end{figure*}

\begin{figure*}[!h]
\centering
\includegraphics[width=\linewidth]{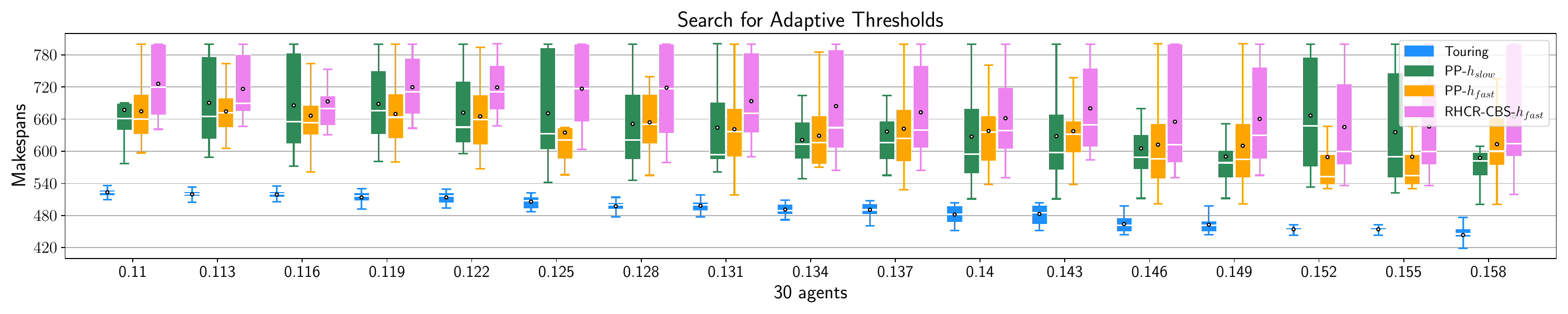}
\includegraphics[width=\linewidth]{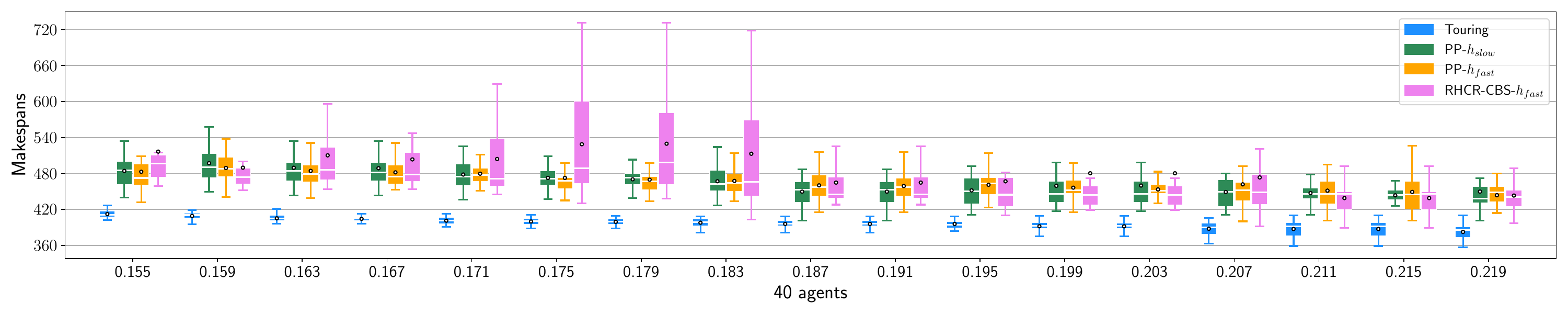}
\includegraphics[width=\linewidth]{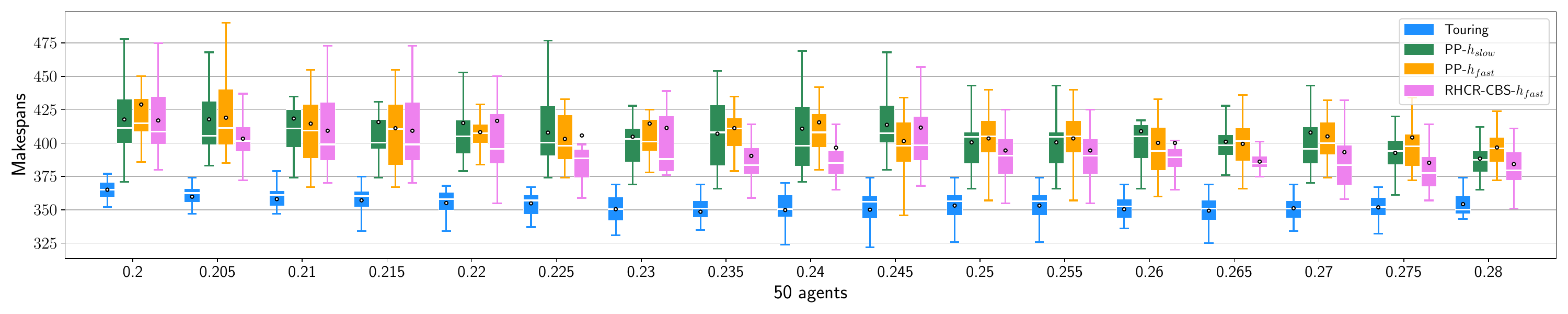}
\includegraphics[width=\linewidth]{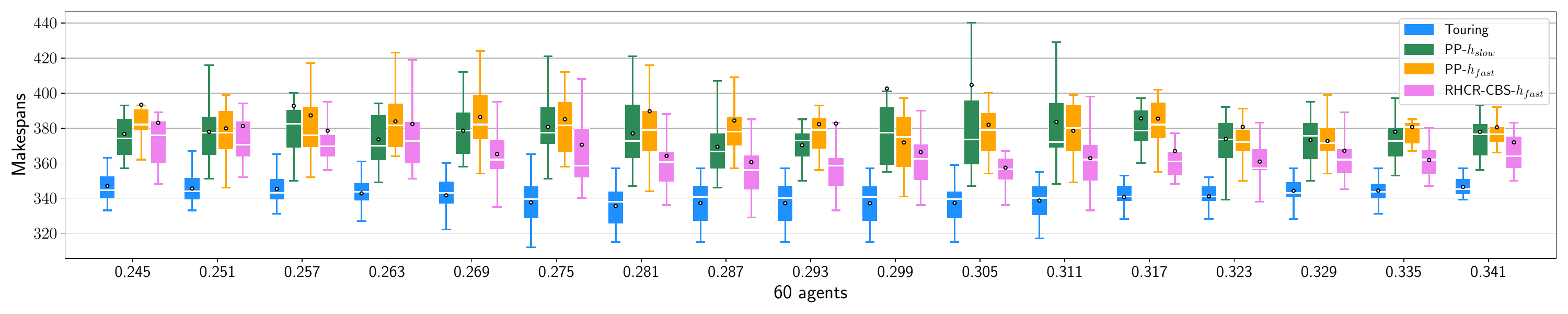}
\includegraphics[width=\linewidth]{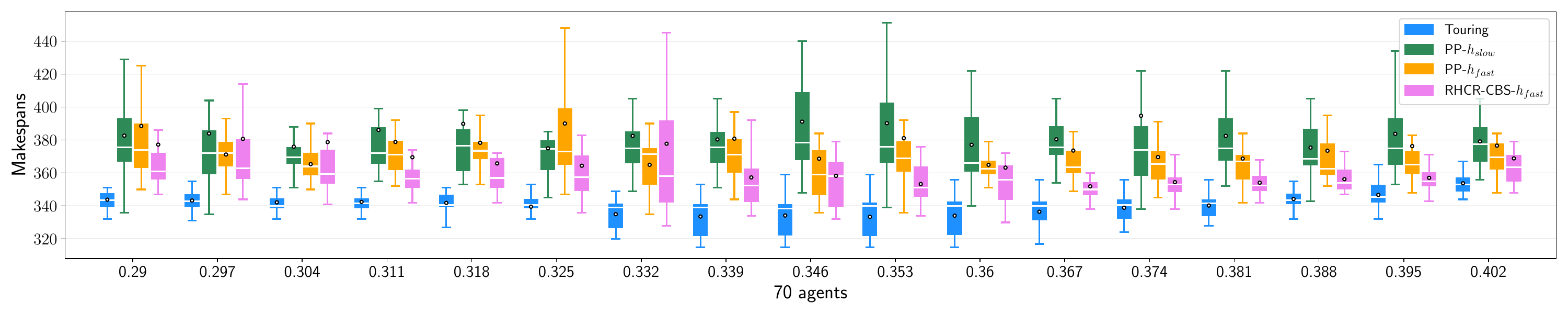}
\caption{The box-plot of makespans over different adaptive thresholds with various scales of agents in Meituan warehouse simulation.}
\label{fig:alpha_bp}
\end{figure*}

In the design of both the \textbf{Touring} and adaptive task assignment, there are certain hyper-parameters.
We here show how the best option is searched in terms of minimizing the eventual makespan.
\begin{enumerate}
	\item Turning frequency (Figure~\ref{fig:turning_freq}). Figure~\ref{fig:eg_non_wf} has presented the extreme where every possible cell that can be a turning is set as a turning, i.e., of frequency 1. One can gradually ``sparsify'' the turnings to see if the overall makespan gets worse. It turns out, the more turnings you have, the better the makespan on average will be. 
	\item Adaptive Threshold (Figure~\ref{fig:alpha_bp}). As the occupation ratio is defined as the number of agents over the number of passable cells in that part of area, the spectrum of tested thresholds in $N$-agent scenarios will be considerably less than those in $N'$-agent scenarios if $N < N'$.
	One can clearly observe that our \textbf{Touring} planner significantly outperforms the other three, and the threshold that makes the lowest box plot is the most desired one. Another observation from Figure~\ref{fig:alpha_bp} is that ours is also more stable than the other three, as the variations (the length of those boxes) are relatively small in most cases.
\end{enumerate}

\section{RL Training Details and Remarks}
\label{apd:rl_train}

Here we reveal the details of RL training skipped in Section~\ref{sec:ta_rl}.

\textbf{Actions.} We directly mask out unavailable actions (those delivery ports that do not need the item) at each assignment state, instead of signaling large negative rewards. In principle, these two are equivalent in terms of the value of the eventual optimal policy, but the former one will guide the policy optimization to converge faster~\cite{huang2022closer}.

\textbf{State features.} As defined in Section~\ref{sec:ta_rl}, assignment states contain necessary information from system-states. Here we make each state of size $num\_of\_agents\times (2 + 1)$,
	which means to mark each agent's location and direction (converted to [0, 90, 80, 270]). The location feature is further normalized by the layout shape, and the direction feature is normalized by 360.
	
\textbf{Episodes.} We train the RL agents over one set of item sequences while evaluate it over another set of item sequences.

\begin{figure*}[!ht]
\centering
\includegraphics[width=0.96\linewidth]{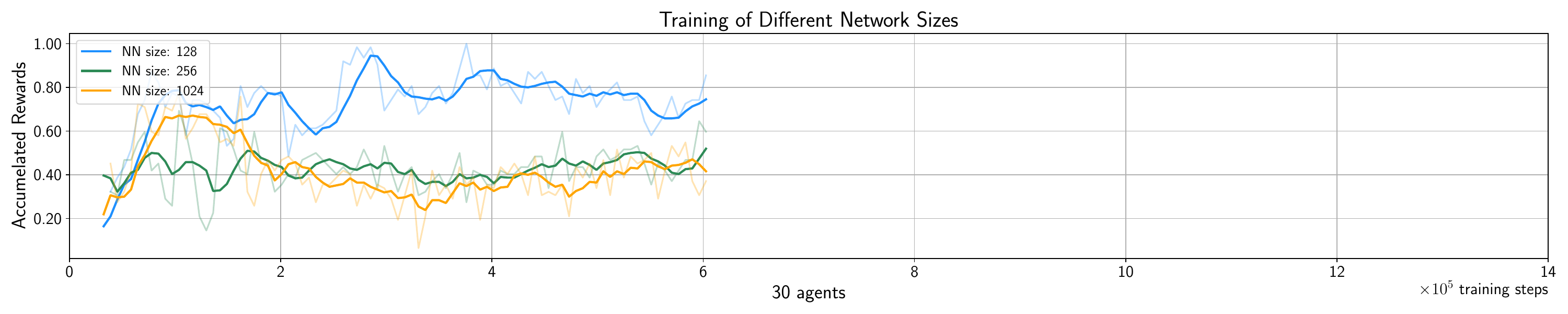}
\includegraphics[width=0.96\linewidth]{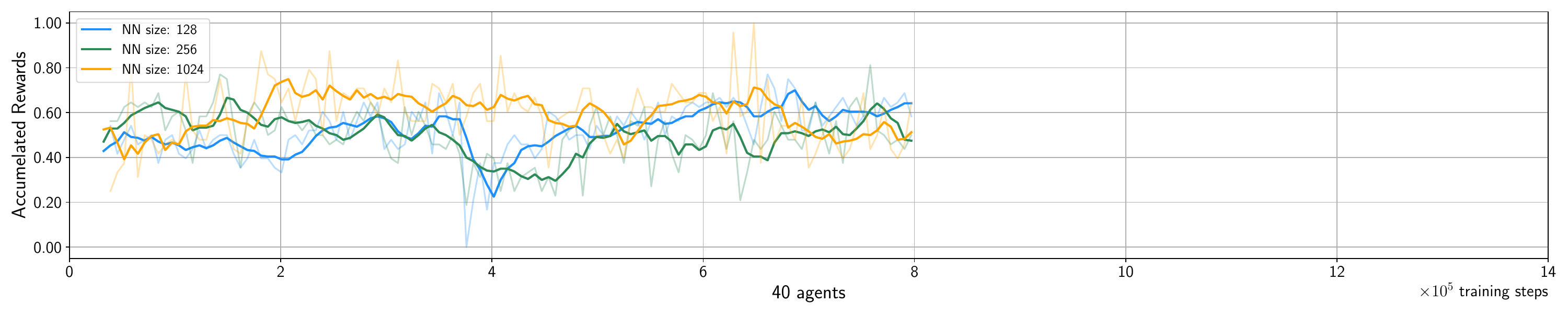}
\includegraphics[width=0.96\linewidth]{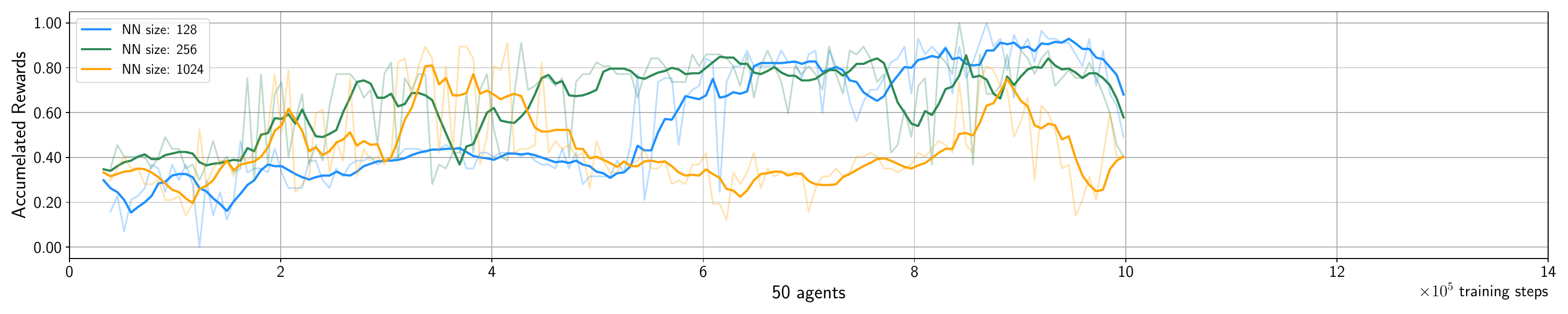}
\includegraphics[width=0.96\linewidth]{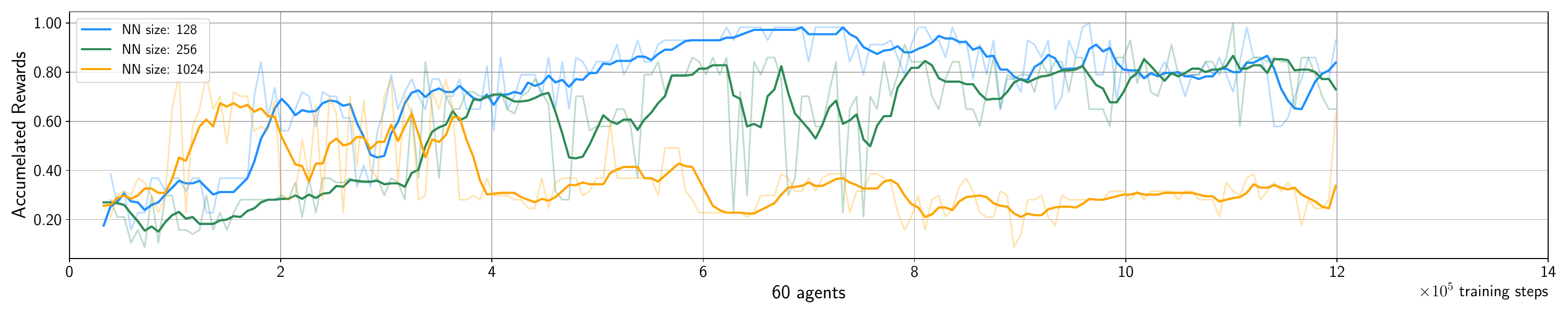}
\includegraphics[width=0.96\linewidth]{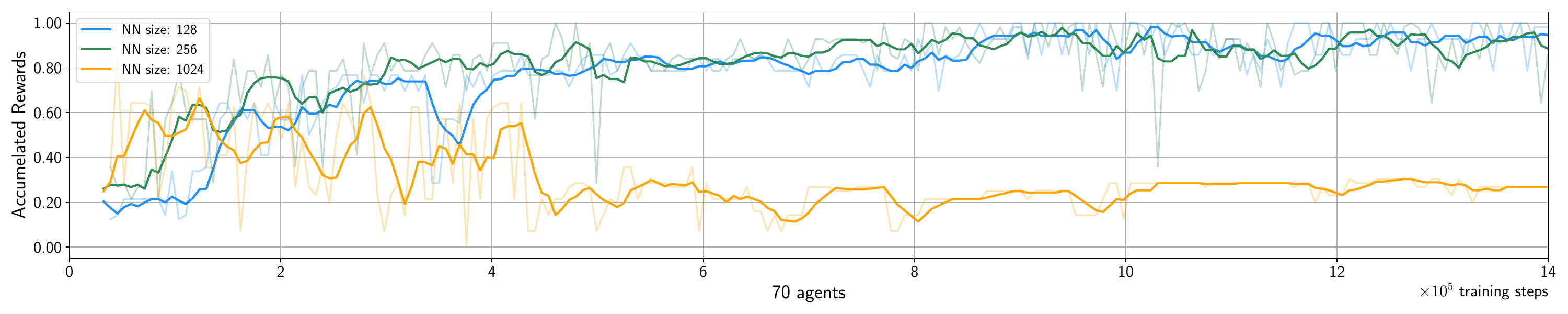}
\caption{Training processes for different network sizes in various scales of agents. }
\label{fig:rl_log}
\end{figure*}

\textbf{Hyper-parameters.} Both the value network and the policy network are MLPs of size $H \times H \times H \times H$ followed by respective value/policy heads.
We attach some training samples in Figure~\ref{fig:rl_log}, for $H$ chosen from [128, 256, 1024]. The number of total training steps can also be seen in this figure. It turns out networks with $H=1024$ tend to overfit in most cases.

%

\textbf{Necessity of RL.} In Table~\ref{tab:eval_full}, RL performs better than adaptive methods with 30/40 agents, though slightly worse with 50/60/70 agents. Please also note that for the column $RL^{best}$, the number$^\text{(in superscripts)}$ is always the best.
RL is used here to solve the underlying MDP, which is a direct characterization of task assignment. This systematic formulation is crucial for a comprehensive study, even though the training outcomes may lag in practice.
Additionally, while the closer-first strategy seems effective in the experiments, such fixed-priority methods are not desirable in practice.
Similar to processor scheduling in OS, a significant drawback is that certain tasks may continually be preempted, i.e., a delivery port gets no item for a long operational period.

\textbf{Delayed rewards.} A critical challenge is mentioned in Section~\ref{sec:ta_rl}, termed \textit{delayed rewards}.
By our definition, the reward is the negative time cost spent between two assignment states. These reward signals are quite ``delayed'', in the sense that for two adjacent assignment states, the immediate reward received at the latter one might not reflect the delivery cost for the task assigned in the former state (it is usually not delivered yet).
This issue is worsened when the number of agents grows, as a longer sequence of transitions is needed for an item to be delivered.

\textbf{Alternative algorithms.} Other than PPO, we have also considered DQN and REINFORCE,
expecting REINFORCE to outperform PPO and DQN, as it collects accumulated rewards for the entire episode in a Monte Carlo manner without learning a value model, instead of minimizing the temporal difference error like PPO and DQN. However, PPO performed the best eventually.

\section{Running Example Recordings}
We provide three video clips under the \texttt{/figs} folder of the online repository\footnote{\url{https://github.com/Fernadoo/Online-TAPF}} for the \textit{Touring} planner coupled with the \textit{closest-first} assigner, the \textit{adaptive} assigner with $\alpha = 0.235$, and the \textit{RL} assigner with the best best-case performance, respectively.
\begin{enumerate}
	\item \texttt{warehouse\_50\_touring\_closest.mp4} with makespan 355.
	\item \texttt{warehouse\_50\_touring\_alpha0235.mp4} with makespan 325.
	\item \texttt{warehouse\_50\_touring\_rl.mp4} with makespan 316.
\end{enumerate}
In all the above instances, the initial states are the same, i.e. the corresponding agents are in the same locations and towards the same directions at the beginning, and the online item sequences are also the same.

\end{document}